\documentclass[11pt]{article}
\usepackage[top=1in,bottom=1in,left=1in,right=1in]{geometry}
\usepackage{tikz}
\usepackage{tikzscale}
\usepackage{balance}
\usepackage{pgfplots}
\usepackage{subfig}
\usepackage{times}
\usepackage{amssymb}
\usepackage{graphicx}
\usepackage{amsmath}
\usepackage{booktabs}
\usepackage[algo2e,linesnumbered,ruled,vlined]{algorithm2e}
\usepackage{algorithm}
\SetCommentSty{textrm}
\usepackage{listings}
\usepackage{color}
\usepackage{url}

\usepackage{hyperref}
\usepackage{euscript}
\usepackage{wrapfig}

\usepackage[skip=3pt]{caption}

\newif\iffull
\fullfalse

\newcommand{\remove}[1]{}

\newcommand{\revcolor}[1]{%
\ifnum#1=0 red\fi%
\ifnum#1=1 blue\fi%
\ifnum#1=2 green\fi%
}

\newcommand{\tightfloats}{
  \setlength{\floatsep}{1ex}
  \setlength{\textfloatsep}{1ex}
  \setlength{\intextsep}{1ex}
  \setlength{\dblfloatsep}{\floatsep}
  \setlength{\dbltextfloatsep}{\textfloatsep}
  \setlength{\abovecaptionskip}{0ex}
  \setlength{\belowcaptionskip}{0ex}
  \renewcommand{\topfraction}{1.00}
  \renewcommand{\bottomfraction}{1.00}
  \renewcommand{\floatpagefraction}{1.00}
  \renewcommand{\dbltopfraction}{1.00}
  \renewcommand{\dblfloatpagefraction}{1.00}
  \renewcommand{\textfraction}{0.0}
}

\makeatletter
\def\enumhook{}

\def\itemhook{}

\def\descripthook{}
\def\enumerate{%
  \ifnum \@enumdepth >\thr@@\@toodeep\else
    \advance\@enumdepth\@ne
    \edef\@enumctr{enum\romannumeral\the\@enumdepth}%
      \expandafter
      \list
        \csname label\@enumctr\endcsname
        {\usecounter\@enumctr\def\makelabel##1{\hss\llap{##1}}%
          \enumhook \csname enumhook\romannumeral\the\@enumdepth\endcsname}%
  \fi}
\def\itemize{%
  \ifnum \@itemdepth >\thr@@\@toodeep\else
    \advance\@itemdepth\@ne
    \edef\@itemitem{labelitem\romannumeral\the\@itemdepth}%
    \expandafter
    \list
      \csname\@itemitem\endcsname
      {\def\makelabel##1{\hss\llap{##1}}%
        \itemhook \csname itemhook\romannumeral\the\@itemdepth\endcsname}%
  \fi}

\makeatother

\newcommand{\denselists}{%
  \renewcommand{\itemhook}{%
    \setlength{\topsep}{0.5ex}%
    \setlength{\leftmargin}{3ex}%
    \setlength{\partopsep}{0ex}%
    \setlength{\parsep}{\parskip}%
    \setlength{\itemsep}{0.5ex}%
  }%
  \renewcommand{\enumhook}{\itemhook}%
  \renewcommand{\descripthook}{\itemhook}%
}

\denselists
\tightfloats
\newtheorem{example}{Example}

\renewcommand{\Re}{{\rm I\!\hspace{-0.025em} R}}

\newcommand{\Expec}[1]{\mathbf{E}\left [ #1 \right ]}

\newcommand{\NP}{\mathrm{NP}}

\newcommand{\polyn}{\mathop {\mathrm{poly}}}

\newtheorem{theorem}{Theorem}[section]
\newtheorem{lemma}[theorem]{Lemma}

\newenvironment{proof}[1]{\trivlist\item[]\emph{Proof#1}:}%
                  {\unskip\nobreak\hskip 1em plus 1fil\nobreak%
                           \rule{2mm}{2mm}
                           \parfillskip=0pt%
                           \endtrivlist}

\def\*#1{\mathbf{#1}}
\newcommand{\Reals}{\ensuremath{\mathbb{R}}}
\newcommand{\Objs}{\ensuremath{\EuScript{O}}}
\newcommand{\obj}{\ensuremath{o}}
\newcommand{\transpose}[1]{\ensuremath{{#1}^\intercal}}
\newcommand{\One}[1]{\ensuremath{\mathbf{1}\left[#1\right]}}
\newcommand{\Prob}[1]{\mathbf{Pr}\left[\,#1\,\right]}
\newcommand{\Expected}[1]{\ensuremath{\mathbf{E}\left[\,#1\,\right]}}
\newcommand{\Var}[1]{\mathbf{Var}\left[\,#1\,\right]}
\newcommand{\Cov}[1]{\mathbf{Cov}\left[\,#1\,\right]}
\newcommand{\card}[1]{\ensuremath{\left|#1\right|}}
\newcommand{\concat}[2]{\ensuremath{\overline{#1,#2}}}
\newcommand{\concatiii}[3]{\ensuremath{\overline{#1,#2,#3}}}

\DeclareMathOperator*{\argmax}{arg\,max}
\renewcommand{\emptyset}{\varnothing}

\newcommand{\bias}[2]{\ensuremath{\mathsf{bias}(#1,#2)}}
\newcommand{\duplicity}[2]{\ensuremath{\mathsf{dup}(#1,#2)}}
\newcommand{\fragility}[2]{\ensuremath{\mathsf{frag}(#1,#2)}}

\newcommand{\MinVar}{\ensuremath{\mathsf{MinVar}}}
\newcommand{\MinVarComplement}{\ensuremath{\overline{\mathsf{MinVar}}}}
\newcommand{\EVar}[1]{\ensuremath{\mathsf{EV}(#1)}}
\newcommand{\EVarComplement}[1]{\ensuremath{\overline{\mathsf{EV}}(#1)}}
\newcommand{\MaxPr}{\ensuremath{\mathsf{MaxPr}}}

\newcommand{\Greedy}{\ensuremath{\mathsf{Greedy}}}
\newcommand{\GreedyNaive}{\ensuremath{\mathsf{GreedyNaive}}}

\newcommand{\OPT}{\ensuremath{\mathrm{OPT}}}

\newcommand{\OptF}{\textbf{Optimum}}
\newcommand{\GreedyF}{\textbf{Greedy}}
\newcommand{\BlindF}{\textbf{GreedyBlindCost}}
\newcommand{\RandomF}{\textbf{Random}}

\newcommand*\diff{\mathop{}\!\mathrm{d}}

\def\mparagraph#1{\par\smallskip\noindent\textbf{#1.}\quad}

\allowdisplaybreaks

\begin{document}

\title{Selecting Data to Clean for Fact Checking: Minimizing Uncertainty vs.\ Maximizing Surprise}

\author{Stavros Sintos\thanks{ssintos@cs.duke.edu, Department of Computer Science, Duke University} \and Pankaj K. Agarwal\thanks{pankaj@cs.duke.edu, Department of Computer Science, Duke University} \and Jun Yang\thanks{junyang@cs.duke.edu, Department of Computer Science, Duke University}}
\date{\vspace{-6ex}}

\pagenumbering{gobble}

\maketitle
\pagenumbering{arabic}

\begin{abstract}
  We study the optimization problem of selecting numerical quantities
  to clean in order to fact-check claims based on such data.
  Oftentimes, such claims are technically correct, but they can still
  mislead for two reasons. First, data may contain uncertainty and
  errors. Second, data can be ``fished'' to advance particular
  positions. In practice, fact-checkers cannot afford to clean all
  data and must choose to clean what ``matters the most'' to checking
  a claim. We explore alternative definitions of what ``matters the
  most'': one is to ascertain claim qualities (by minimizing
  uncertainty in these measures), while an alternative is just to
  counter the claim (by maximizing the probability of finding a
  counterargument). We show whether the two objectives align with
  each other, with important implications on when fact-checkers should
  exercise care in selective data cleaning, to avoid potential bias
  introduced by their desire to counter claims. We develop efficient
  algorithms for solving the various variants of the optimization
  problem, showing significant improvements over naive solutions. The
  problem is particularly challenging because the objectives in the
  fact-checking context are complex, non-linear functions over
  data. We obtain results that generalize to a large class of
  functions, with potential applications beyond fact-checking.
\end{abstract}

\section{Introduction}
\label{sec:intro}

We proud ourselves in basing our decisions on data and evidence, yet
``data determinism'' is not without its own issues.  Two glaring
issues are data quality---where inaccurate data leads to incorrect
conclusions---and the practice of \emph{data fishing}---where, even
when assuming perfect data, one can cherry-pick data to make correct
but still misleading claims.  Journalists and fact-checkers devote an
enormous amount of effort to checking claims based on data, and have
to struggle constantly with both issues, often with limited time and
resources.  Consider the following example.

\begin{example}\label{exmp:ex1}
  Our world is never short of controversial and contradictory claims
  about crime statistics.  \emph{``There have been huge drops in the
    murder rates in cities.''}  \emph{``Neighborhoods have become more
    violent under his watch.''}  The list goes on.  While we have left
  out the specifics here, a few internet searches will reveal a long
  list of such claims, all seemingly backed by data.

  Fact-checking such claims is no easy task.  First, there are
  numerous well-documented data quality issues with crime
  data~\cite{Maltz-99-gaps_crime_data, FiveThirtyEight-Bialik-16}.  In
  the U.S., a primary source for such data is the \emph{Uniform Crime
    Reports}, which relies on voluntary reporting from law enforcement
  agencies at the different levels of jurisdiction.  Coding errors and
  inconsistencies, changes to reporting guidelines, or even reporting
  delays and personnel changes can lead to under- or overstatement of
  crimes for particular jurisdictions in particular time periods.  To
  ``clean'' a potentially erroneous value, a fact-checker may need to
  go through local agencies and/or consult with experts who have
  previously worked with the data.  With limited time and resources,
  it is often impractical to clean all relevant data items.

  Second, many claims are correct but misleading.  For example, a
  claim about rising or falling crime rate may be made over a
  particular time period.  However, if we simply change the period a
  bit, the trend may become less pronounced or even reversed.  In that
  case, the claim is not fair or robust, and we can show a ``counter''
  claim, similar in form to the original but with a different
  conclusion, to help refute the original.  As another example, if the
  claim is intended as an attack on the leadership for a particular
  jurisdiction, we can check whether similar claims can be made for
  other jurisdictions or for other previous leaderships.  If yes, then
  the original claim is not unique.  Fact-checkers frequently employ
  such analyses in rebutting correct but misleading claims,
  e.g.~\cite{FiveThirtyEight-Bialik-16, Gathright-16-darry_glenn,
    Robertson-16-dueling_crime_claims}.
\end{example}

To combat the problems of data quality and data fishing, we can draw
methods from data cleaning~\cite{DBLP:journals/ftdb/IlyasC15,
  DBLP:series/synthesis/2013Ganti} and recent work on fact-checking
correct but misleading claims~\cite{wu2014toward,
  tods17-WuEtAl-comp_fact_check}.  This paper explores the following
specific question: given a limited budget to fact-check a claim, which
data items should we choose to clean?  Intuitively, we would like to
prioritize efforts towards those parts of data that ``matter the
most'' to fact-checking the given claim.

Consider claims that can be modeled as queries over a database.  When
the values in the database are uncertain, the correctness of the claim
(query result) is uncertain too.  We can spend some budget to remove
uncertainty in some data values, which can help reduce uncertainty in
claim correctness.  For fact-checking, however, we must go beyond
correctness.  Instead, following the perturbation framework
of~\cite{wu2014toward, tods17-WuEtAl-comp_fact_check}, we consider
\emph{perturbations} of the original claim, which provide a larger
context in which we can assess various claim qualities---including
\emph{fairness}, \emph{robustness}, and \emph{uniqueness}---or to find
counterarguments.  The goal of data cleaning hence needs to be
extended to help with such analyses.

An important question is how the objective of cleaning (i.e., how we
define what ``matter the most'') affects fact-checking.  A reasonable
objective is minimizing uncertainty in some numeric measure of claim
quality (e.g., fairness, uniqueness)---the goal is to ascertain claim
quality.  Another possibility is maximizing the chance of finding a
counterargument to the given claim after cleaning---the goal is to
purely counter the claim.  One key question is whether these two goals
align with each other.  If not, fact-checkers need to be careful in
avoiding potential bias of their data cleaning choices introduced by
their desire to counter claims.

\begin{example}\label{eg:concrete}
  To illustrate, consider the numbers of crimes in a
    particular jurisdiction in recent years (subscripts are
    years):\\\centerline{\small%
    \begin{tabular}{c|c|c|c|c}
      $X_{2014}$ & $X_{2015}$ & $X_{2016}$ & $X_{2017}$ & $X_{2018}$\\\hline
      $9{,}010$ & $9{,}275$ & $9{,}300$ & $9{,}125$ & $9{,}430$
    \end{tabular}%
  } Suppose the data may contain errors.  Cleaning each $X_i$ may
  yield a number different from the above, but it would take
  considerable effort and we do not have enough resources to clean
  every $X_i$.

We wish to check the claim ``crimes (in 2018) have gone up
  by more than $300$ cases from last year,'' which attempts to put the
  blame on the current administration.  This claim can be modeled as a
  simple query $X_{2018}-X_{2017}$.  Obviously, cleaning $X_{2018}$
  and $X_{2017}$ will let us tell whether the claim is outright
  incorrect.

But fact-checking goes beyond verifying correctness.
  Implicitly, this claim suggests that having an increase of $300$ in
  a year is an unusual event.  To assess whether this claim is really
  ``unique,'' we consider a series of perturbations (more in
  Section~\ref{sec:model:check}), i.e., additional queries that help
  put the original one in context---how much did crimes go up by from
  2016 to 2017, from 2015 to 2016, and from 2014 to 2015?  We can then
  quantify the uniqueness of the original claim by counting the number
  of perturbations that yield a result no weaker than the original
  (i.e., $> 300$).  To make this assessment, we would need a whole lot
  of data beyond the two specific years originally referenced.

The question of what data to clean now becomes more
  involved.  Suppose our goal is to purely counter the claim by
  finding another instance of big increase in recent years.  For this
  simple example, we would intuitively want to clean $X_{2015}$.  If
  the result goes up just by a little, say, from $9{,}275$ to
  $9{,}315$, we will be able to make the counterargument that crimes
  went up just as much from 2014 to 2015 (implying that the previous
  administration could be blamed too).  In contrast, cleaning
  $X_{2016}$ is probably not a good investment, because it will
  unlikely yield a high enough number to make the increase over
  2015--2016 significant, or a low enough number to make the increase
  over 2016--2017 significant.

On the other hand, the cleaning strategy above begs the
  question of whether it is ``cherry-picking'' in a way that can lead
  to unfair assessment of the original claim.  A more impartial
  objective would be minimizing uncertainty in some measure of claim
  quality (e.g., uniqueness here).  Would this objective lead to very
  different choices of data items to clean?  Are there situations
  where two objectives actually align with each other?  These are some
  of the questions this paper seeks to answer.
\end{example}

Note that this paper focuses on how to select data to clean, as
opposed to specific data cleaning techniques.  We assume that a
cleaning procedure can resolve the uncertainty in a value by paying a
cost; our algorithms are general enough to work for multiple
scenarios, including manual cleaning.

\mparagraph{Our contributions}
Given a claim to check against a database with uncertain values, we
solve the problem of choosing what values to clean under a cost budget
in order to---roughly speaking---either 1) minimize uncertainty in
some measure of claim quality, or 2) maximize the chance of finding a
counterargument after cleaning.  By appropriately defining a query
function $f$ over the database, we show that the above problem reduces
to the following, more fundamental problems of choosing data to clean
under a budget constraint, with different objectives:
\begin{itemize}
\item (\MinVar) Minimize the uncertainty in the result of $f$, or
  more precisely, the expected variance in the result of $f$ after
  cleaning.
\item (\MaxPr) Maximize the probability that the result of $f$ after
  cleaning deviates significantly from its result before.
\end{itemize}
These optimization problems in general have applications beyond
fact-checking.  We give hardness results and greedy algorithms that
work well in practice.  Under certain assumptions, we can exploit
properties of the data distribution and query function $f$ to obtain
efficient algorithms with good approximation
guarantees.\footnote{\small We note that our solution to \MinVar\ is
  also of technical interest because of its connection to the
  submodular function maximization problem with applications in sensor
  placement to reduce uncertainty~\cite{krause2008near,
    krause2008robust}.  Despite apparent similarity, there is an
  intriguing dichotomy between the two problems that necessitates
  different approaches.  We point out this dichotomy in Section~\ref{sec:related}.}

We apply our results to fact-checking and evaluate our algorithms
using experiments.  Fact-checking poses hard instances of the above
problems (e.g., query functions can involve indicator and quadratic
functions), but our results are general enough to apply.  For the
question regarding the differing goals of fact-checking, we show that
in general, these two goals do not align.  More interestingly, we also
show that, under certain assumptions that may be reasonable in
practice, these two goals can in fact align with each other.  These
results provide practical tools and guidelines that help fact-checkers
clean data effectively while avoiding potential bias.


\section{Model}
\label{sec:model}

We will first define our problems generally in terms of a query
function $f$ over uncertain data.  The objective is to clean some data
to either minimize the uncertainty in the result of $f$, or maximize
the probability that the result of $f$ after cleaning deviates
significantly from the result before.  Then, we show how to map
concrete fact-checking tasks to special instances of these general
problems, by defining appropriate $f$ and choosing an objective.

\subsection{Problem Definition}
\label{sec:model:problem}

Let $\Objs = (\obj_1, \ldots, \obj_n)$ be a set of $n$ objects.  We
assume that their identities are certain but their \emph{values} are
not.  Each object $\obj_i$ has a \emph{current value} $u_i\in \Re$,
which may be incorrect; let $\*u = \transpose{(u_1, \ldots, u_n)}$. We
model the \emph{true value} of $\obj_i$ as a random variable $X_i$
with support $V_i$, and assume that we know the joint probability
distribution of $\*X = \transpose{(X_1, \ldots, X_n)}$ with support
$\*V$. Cleaning object $\obj_i$ costs $c_i$, and reveals its true
value drawn from $\*X$.

A \emph{query function} $f$ is a real-valued function of the object
values. Informally, given $f$ and a cost budget $C$, our problem is
to choose a subset of objects $T \subseteq \Objs$ to clean in order to
1)~minimize the uncertainty in $f(\*X)$, or 2)~maximize the
probability that cleaning $T$ leads to a large deviation from
$f(\*u)$. We describe the two objectives further below and formally
define the problems.

Given
$T = \{ \obj_{i_1}, \obj_{i_2}, \ldots, \obj_{i_{\card{T}}} \}
\subseteq \Objs$
where $1 \le i_1 < i_2 < \cdots < i_{\card{T}} \le n$, we denote
$\transpose{(X_{i_1}, X_{i_2}, \ldots, X_{i_{\card{T}})})}$ by $\*X_T$
and its support by $\*V_T$. In general, by subscripting an $n$-vector
with $T = \{ \obj_{i_1}, \ldots, \obj_{i_{\card{T}}} \}$, we mean the
$\card{T}$-vector consisting of just those elements at positions
$i_1, \ldots, i_{\card{T}}$.

\mparagraph{Minimizing uncertainty}
The outcome of cleaning $T$ is uncertain. While cleaning a particular
$\obj_i$ always removes uncertainty in $X_i$, doing so may,
counterintuitively, increase uncertainty in $f(\*X)$, as illustrated
by the following example.

\begin{example}[Uncertain effect of cleaning]%
  \label{eg:uncertainty-increase}
  Consider a database of three objects with binary values $X_1$,
  $X_2$, and $X_3$, and query function
  $f(\*X) = \One{X_1 + X_2 + X_3 < 3}$, an indicator function that
  returns $1$ if the sum of values is less than $3$, or $0$ otherwise.
  Intuitively, the indicator condition may become harder or easier to
  satisfy depending on the outcome of cleaning an object value, say
  $X_1$. As a concrete example, suppose $X_1$, $X_2$, and $X_3$ are
  independent Bernoulli random variables with success probabilities
  $1 \over 2$, $1 \over 3$, and $1 \over 4$, resp. Without cleaning,
  $f(\*X) = 0$ with probability
  ${1 \over 2} \cdot {1 \over 3} \cdot {1 \over 4} = {1 \over 24}$.
  If we clean $X_1$, there are two outcomes:
  If $X_1 = 0$: here we know the sum cannot exceed $2$, so
    $f(\*X) = 1$ for sure, and uncertainty is reduced.
  Otherwise, $X_1 = 1$: in this case, $f(\*X) = 0$ with probability
    $\Prob{X_2+X_3 \ge 2} = {1 \over 3} \cdot {1 \over 4} = {1 \over
      12}$,
    which is closer to a toss-up than the probability of $1 \over 24$
    for the case without any cleaning. In other words, uncertainty
    has increased.
\end{example}

Hence, when choosing what to clean, we can only minimize uncertainty
in the expected sense. To this end, consider $\*V_T$, which forms the
sample space containing all possible outcomes of cleaning $T$. The
uncertainty in the query function result (due to remaining uncertainty
in $\Objs \setminus T$) can be regarded as a random variable over
$\*V_T$. Our objective is then to minimize its expected value.

There are various measures of uncertainty in random variables
(e.g. entropy). In this paper we consider variance, which is useful
when the actual spread and magnitude of the numerical quantity matters
(for example the number of crimes in Example \ref{exmp:ex1}).
Formally, we define the optimization problem \emph{\MinVar} as
follows:
\begin{equation}
  \begin{split}
    \textbf{(\MinVar)}\;%
    & \min_{T\subseteq \Objs}\sum_{\*v \in \*V_T}\Prob{\*X_T = \*v}
    \cdot \rlap{$\Var{f(\*X) \mid \*X_T = \*v},$}\hspace*{8em}\mbox{}\\
   \text{subject to:}\;%
    & \sum_{\obj_i \in T} c_i \le C.
  \end{split}
\end{equation}
Note that the above definition assumes the value domains to be
discrete; the case when they are continuous can be defined analogously
by integrating a probability density function.

Notice that so far, the current object values ($\*u$)
have no bearing on the problem definition. However, in the scenario
to be discussed next, the current values play a more important role.

\mparagraph{Maximizing surprise}
The value of $f$ before cleaning $T$ is $f(\*u)$. By cleaning $T$, we
replace, for each object $\obj_i \in T$, the current value $u_i$ with
a random draw of $X_i$, with the hope of lowering the value of $f$ by
more than a given threshold $\tau \ge 0$---intuitively, finding a
``surprise.'' Because the outcomes are random, we maximize the
probability that our goal is met. Formally, the optimization problem,
which we call \emph{\MaxPr}, is as follows:
\begin{equation}
  \begin{split}
    \textbf{(\MaxPr)}\;%
   & \max_{T \subseteq \Objs}\Prob{f(\*X) < f(\*u) - \tau \mid \*X_{\Objs \setminus T} = \*u_{\Objs \setminus T}},\\
     \text{subject to:}\;%
    & \sum_{\obj_i \in T} c_i \le C.
  \end{split}
\end{equation}
Note the objective function has value $0$ for $T = \emptyset$.

\mparagraph{Discussion}
Note that we assume each object $o_i$ has a cleaning cost $c_i$; i.e.,
if we want to learn its correct value we need to pay $c_i$.  This cost
model has been widely used in data cleaning literature
(e.g.,~\cite{chen2008quality, cheng2008cleaning,
  dallachiesa2013nadeef, liu2005cost}).  It is general enough to allow
any individual costs to be specified, be they estimated or actual,
monetary costs or human efforts.  However, we do not yet consider more
general cases where the cost function is non-additive.

We also assume that the distribution for each object's value is known.
Such distributional knowledge can come in many ways, e.g., from the
design of some sampling procedure, by modeling measurement errors of
sensors, by resolving conflicting data from different
sources~\cite{dong2009integrating}, or by \emph{opinion
  pooling}~\cite{chang1986combination, french1980updating,
  kleinberg1999authoritative}.  Such knowledge has been used in other
related data cleaning models (e.g.,~\cite{cheng2008cleaning}) as well
as probabilistic databases (e.g.,~\cite{cavallo1987theory,
  dalvi2009probabilistic, dalvi2007efficient,
  suciu2011probabilistic}).

\subsection{Application in Fact Checking}
\label{sec:model:check}

We now show how to apply the general problems defined earlier to
concrete fact-checking tasks.
Following~\cite{wu2014toward, tods17-WuEtAl-comp_fact_check}, we model
a claim as a query (called the \emph{claim function}) over a database
instance. Let $q^\circ$ denote the claim function for the ``original''
claim to be checked, which returns $q^\circ(\*u)$ given the current
values of the database objects. Intuitively, $q^\circ$ captures the
original claim's particular view of the data. Checking this claim
involves considering various \emph{perturbations} to $q^\circ$ and see
how they compare with $q^\circ(\*u)$. Let $Q = \{ q_1, \ldots, q_m \}$
denote the set of $m$ \emph{perturbations}, each of which is a claim
function obtained by changing (the parameters and/or form of)
$q^\circ$ in some way.

A real-valued \emph{relative strength function} $\Delta(\cdot, \cdot)$
is used to compare two claim function results: if
$\Delta(q_k(\*u), q^\circ(\*u))$ is positive (negative), then $q_k$
strengthens (weakens, resp.) $q^\circ$; the absolute value of
$\Delta(q_k(\*u), q^\circ(\*u))$ measures the extent of strengthening
(weakening, resp.).

Not all perturbations are equally relevant to the original claim. For
example, a perturbation $q_k$ whose parameters are close to those of
$q^\circ$ is more relevant than one whose parameters are far away.
Therefore, we associate each perturbation $q_k$ with a
\emph{sensibility} $s_k \ge 0$ such that
$\sum_{1 \le k \le m} s_k = 1$. The higher the sensibility, the more
relevant this perturbation is to $q^\circ$. Together, the
sensibilities $\mathbf{s} = (s_1, \ldots, s_m)$ define a probability
distribution over $Q$.

\begin{example}[Window aggregate comparison claims]%
  \label{eg:window-aggr-compare}
  A \emph{window aggregate comparison}~\cite{wu2014toward} claim
  compares the aggregate values computed over two time windows of equal
  length.  A real-life example is a claim made by Rudy Giuliani in
  2007~\cite{Jackson-07-nyc_adoption}, which stated that ``adoptions
  went up 65 to 70 percent'' in New York City between the periods
  1990--1995 and 1996--2001.  Here, the claim function is a linear
  function over subsets of values in the windows compared; i.e.,
  $q^\circ(\*u)=\sum_{i=l}^{l+w-1}u_i -\sum_{i=r}^{r+w-1}u_i$, for
  $1\leq l, r\leq n$, where $w$ is length of each window and $u_i$ is the number of adoptions in year $i$.  The first
  summation is over values in the earlier window, while the second
  summation is over the later window.  The $\Delta$ function in this
  case is simply the difference between $q_k(\*u)$ (perturbation) and
  $q^\circ(\*u)$ (original claim).  Sensibility of a perturbation
  $q_k$ in this case may be defined to decay exponentially over its
  distance to $q^\circ$, as measured by the number of years between
  the endpoints of their comparison periods.  The intuition is that we
  care mostly about perturbations with temporal contexts similar to
  the original claim, which is when Guiliani was the mayor.
\end{example}

\mparagraph{Ascertaining claim quality}
The following measures of claim quality were introduced
in~\cite{wu2014toward}. They all involve summarizing, in some way,
over all perturbations, the difference between the result of each
perturbation and that of the original claim function. In the
following, $\*X = \*u$ if there is no uncertainty in the object
values.
\begin{itemize}
\item \emph{\bfseries Fairness} can be measured by the amount of
  \emph{bias} in $q^\circ(\*u)$, defined as
  $
    \bias{q^\circ(\*u)}{\*X} = %
    \sum_{1 \le k \le m} %
    s_k \cdot \Delta(q_k(\*X), q^\circ(\*u)).
 $
  Intuitively, bias of $0$ means perturbations on average return the
  same result as the original claim, so the original claim is fair.
  Negative bias means the original claim exaggerates, while positive
  bias means it understates.
\item \emph{\bfseries Uniqueness} can be measured by the degree of
  \emph{duplicity} in $q^\circ(\*u)$, defined as
  $
    \duplicity{q^\circ(\*u)}{\*X} =\\
    \sum_{1 \le k \le m}
    \ensuremath{\mathbf{1}[\Delta(q_k(\*X),
    q^\circ(\*u)) \ge 0]}.
  $
  Intuitively, duplicity is the number of perturbations that yield
  stronger results than the original claim. The lower the duplicity,
  the more unique the claim.
\item \emph{\bfseries Robustness} can be measured by the
  \emph{fragility} in $q^\circ(\*u)$, defined as
 $
    \fragility{q^\circ(\*u)}{\*X} = %
    \sum_{1 \le k \le m} s_k \cdot (\min\{\Delta(q_k(\*X),
    q^\circ(\*u)), 0\})^2.
 $
  Intuitively, low fragility implies that the original claim is
  robust; i.e., it is difficult to find perturbations that weaken the
  original claim.
\end{itemize}

When the object values $\*X$ are uncertain, the results of claim
functions are uncertain, so the claim quality measures become random
variables over $\*X$, whose uncertainty can be measured by their
variance. A reasonable goal for a fact-checker, given a limited
budget for data cleaning, would be to clean a subset of the object
values in order to minimize the variance in some measure of claim
quality. Because the outcome of data cleaning is uncertain, we
minimize the expected variance over all possible outcomes. This
problem is hence an instance of \MinVar\ introduced in
Section~\ref{sec:model:problem}, with query function $f$ set to the
corresponding claim quality measure.

\mparagraph{Increasing the chance of finding counterarguments}
Consider a fact-checker with a ``random'' strategy, who picks a
perturbation at random---with probability proportional to its
sensibility---and hopes that it weakens the original claim $q^\circ$.
Given the current object values $\*u$, $\bias{q^\circ(\*u)}{\*u}$, the
bias in $q^\circ(\*u)$ as defined in Fairness, computes the
expected extent to which the original claim will be weakened by a
random perturbation. Intuitively, if the bias is well below some
(negative) threshold, then we have a good chance of finding a strong
counterargument to the original claim.

With data uncertainty, we can choose to clean a subset of the object
values, and arrive at a new database instance $\*u'$ consisting of the
resulting values as well as old values from $\*u$ for any objects not
cleaned. The bias in $q^\circ(\*u)$ computed on this new database
instance, $\bias{q^\circ(\*u)}{\*u'}$, would reflect how easy it is to
find a strong counterargument after cleaning. Because the outcome of
cleaning is uncertain, the amount of improvement---between
$\bias{q^\circ(\*u)}{\*u'}$ and the baseline of
$\bias{q^\circ(\*u)}{\*u}$---is uncertain. If our goal is purely to
counter $q^\circ$, we would like to choose objects to clean in a way
to maximize the probability that the improvement is tangible, i.e.,
$\bias{q^\circ(\*u)}{\*u'} < \bias{q^\circ(\*u)}{\*u} - \tau$, where
$\tau \ge 0$ is a user-define threshold. This problem is hence an
instance of \MaxPr\ in Section~\ref{sec:model:problem}
with query function $f(\*X) = \bias{q^\circ(\*u)}{\*X}$.

\mparagraph{Comparing the two objectives}
We now return to the interesting question raised at the beginning of
this paper: how do the two objectives above---ascertaining claim
quality and increasing the chance of finding
counterarguments---compare with each other?

Under certain assumptions the two objectives do agree---namely, if
$\*X$ is a multivariate normal distribution centered at the current
values $\*u$ (independence assumption not needed), then for linear
claim functions, maximizing the chance of finding a counterargument is
equivalent to reducing the uncertainty in fairness. We formally state
the result in the end of Section~\ref{sec:algos}.
Note that the assumption above is not unreasonable in practice. Oftentimes we
have limited prior knowledge of the distribution of $\*X$, and there
is no reason to believe that database values are biased. In such
scenarios, a multivariate normal with current values at the center would
indeed be a reasonable starting assumption, and fact-checkers can rest
assured that the goal of finding counters is equivalent to that of
developing a better understanding of the fairness of the original
claim.

However, the two objectives do not generally agree.
Next, we show how they disagree on a
simple concrete example involving only independently and uniformly
distributed values and in Section~\ref{sec:expr}, we
empirically evaluate how much these objectives diverge.

\begin{example}[Differing fact-checking objectives]%
  \label{eg:diff-objective}
  We consider a database of two objects with values
  $\*X = \transpose{(X_1, X_2)}$, where $X_1$ and $X_2$ are
  independently and uniformly distributed over
  $\{0, {1\over2}, 1, {3\over2}, 2\}$ and
  $\{{1\over3}, 1, {5\over3}\}$, resp. The current (uncleaned) values are
  $\*u = \transpose{(1,1)}$. Note that
  $\Var{X_1} = {2\over5}({1^2+({1\over2})^2}) = {1\over2}$ and
  $\Var{X_2} = {2\over3} \cdot ({2\over3})^2 = {8\over27}$.

  The claim function to be checked is $q^\circ(\*X) = X_1 + X_2$.
  Suppose the only relevant ``perturbation'' of $q^\circ$ is itself
  (i.e., $Q = \{q^\circ\}$), so
  $\bias{q^\circ(\*u)}{\*X} = X_1 + X_2 - q^\circ(\*u) = X_1 + X_2 -
  2$.
  We are given enough budget to clean either $X_1$ or $X_2$, but not
  both. We compare the objective of reducing the expected variance in
  $\bias{q^\circ}{\*X}$ versus that of increasing the chance of
  finding a strong counterargument, where $X_1 + X_2 < {17\over12}$
  (below the baseline of $q^\circ(\*u) = 2$).

  For the first objective, note that with no cleaning at all,
  $\Var{\bias{q^\circ(\*u)}{\*X}} = \Var{X_1} + \Var{X_2} = {1\over2}
  + {8\over27}$; cleaning $X_1$ reduces it to $8\over27$ while
  cleaning $X_2$ reduces it to $1\over2$. Hence, the optimal choice
  (minimizing uncertainty) is to clean $X_1$.

  For the second objective, if we clean $X_1$ (and leave $X_2=1$), we
  have
  $\Prob{X_1 + X_2 < {17\over12}} = \Prob{X_1 < {5\over12}} =
  {1\over5}$;
  if we clean $X_2$ (and leave $X_1=1$), we have
  $\Prob{X_1 + X_2 < {17\over12}} = \Prob{X_2 < {5\over12}} =
  {1\over3}$.
  Therefore, the optimal choice (which maximizes the chance of finding
  counterarguments) is to clean $X_2$.
\end{example}

\section{Algorithms}
\label{sec:algos}

The problems in Section~\ref{sec:model:problem} are difficult in
general---even simpler forms are NP-hard (for example, the well-known
Knapsack problem can be easily reduced to a special instance of the
\MinVar). We begin in Section~\ref{sec:algos:greedy} with
  simple greedy algorithms that prioritize objects to clean until the
  cost budget is exceeded.  These algorithms are used as general
  heuristics for our problems.

Then, in Sections~\ref{sec:algos:modular}
  and~\ref{sec:algos:submodular} we consider certain properties of the
  data distributions and/or query functions that enable better
  theoretical bounds for greedy algorithms or more sophisticated
  algorithms.  More specifically, by assuming independent errors, we
  can show a number of interesting results: a)~With an affine query
  function $f$, the optimization objective of \MinVar\ and that of
  \MaxPr\ (additionally assuming zero-mean errors) become
  modularizable, allowing us to map these problems to Knapsack, for
  which better algorithms exist and even greedy offers constant-factor
  approximation. b)~Remarkably, for any query function $f$, \MinVar\
  is equivalent to minimizing a submodular function with a cost upper
  bound constraint, allowing us to apply existing approximation
  algorithms with theoretical guarantees.

Finally, Section~\ref{sec:algos:check} returns to
  applications in fact-checking.  We discuss how to compute expected
  variance (needed for running our algorithms) for fact-checking
  efficiently.  We also show an interesting coincidence: under some
  conditions, namely when claim are linear and data errors are
  zero-mean multivariate normal, then the objective of minimizing
  uncertainty in claim quality (\MinVar) and that of maximizing the
  chance of finding counters (\MaxPr) in fact align with each other.

\subsection{Greedy Algorithms}
\label{sec:algos:greedy}

\begin{algorithm2e}[t]
  $T \gets \emptyset$; $c \gets 0$\;
  \While{$\exists \obj_i \in \Objs \setminus T: c + c_i \le C$}{
    $\obj_i \gets \argmax_{\obj_i \in \Objs \setminus T:\, c + c_i \le
      C} \beta(\obj_i)/c_i$\nllabel{algo:greedy:select}\;
    $T \gets T \cup \{ \obj_i \}$; $c \gets c + c_i$\;
  }
  \If(\tcp*[f]{check to ensure $2$-approximation})%
  {$\Objs \setminus T \neq \emptyset$}{
    $\obj_i \gets \argmax_{\obj_i \in \Objs \setminus T:\, c_i \le C}
    \beta(\obj_i)/c_i$\;
    \If{$\beta(\obj_i) > \sum_{\obj_j\in T}\beta(o_j)$}{
      $T \gets \{\obj_i\}$\;
    }
  }
  \Return{$T$}\;
  \caption{\label{algo:greedy}\Greedy. The algorithm is parameterized by
    $\beta: \Objs \to \Reals$, a function that returns the estimated
    benefit of cleaning a given object. In general, $\beta$ may refer
    to the set $T$ of objects that have already been chosen.}
\end{algorithm2e}

Algorithm~\ref{algo:greedy} shows the template for our greedy
algorithms. The code is parameterized by a \emph{benefit estimation
  function} $\beta: \Objs \to \Reals$, to score objects for greedy
selection. Intuitively, \Greedy\ chooses the object with the highest
benefit per unit cost ($\beta(\obj_i)/c_i$) to clean next until the
total cost exceeds $C$. In the end, we ensure that we always find a
good solution, in particular a $2$-approximation, for a class of
objective functions~\cite{ibarra1975fast} (special cases of our
defined problems) by checking if the next object $\obj_l$ with highest
ratio $\beta(\obj_l)/c_l$ has larger benefit than the sum of benefits
of the objects we have chosen before.  For example, consider the well
known 0-1 knapsack problem with two items $x_1, x_2$.  The values of
the items are $\beta(x_1)=0.1$ and $\beta(x_2)=10$, while the costs
are $c_1=0.0001$ and $c_2=2$.  With budget $C=2$, the goal is to
choose a set of items with cost at most $2$ with the maximum value.
\Greedy\ would choose item $x_1$, because
$\frac{0.1}{0.0001}>\frac{10}{2}$, and hence the value of the knapsack
is $0.1$.  However, the optimal choice is to select item $x_2$ with
value $10$. In the end, our algorithm considers the benefit of the
next non-cleaned object (Lines 5-8 in Algorithm~\ref{algo:greedy}) so
it ensures that in such a case we take the item $x_2$ in our final
solution.

\mparagraph{A naive greedy algorithm}
The simplest instance of \Greedy, which we call \GreedyNaive, uses the
benefit estimation function $\beta(\obj_i) = \Var{X_i}$ (but $0$ if
the query function does not reference $X_i$). Note that
$\beta(\cdot)$ does not refer to the objects already chosen, so one
can sort $\Objs$ by $\beta(\obj_i)/c_i$ once and then proceed
accordingly, without computing the maximum ratio in each iteration.
Therefore, the running time of \GreedyNaive\ is $O(n(t+\log n))$,
where $t$ is the complexity of computing each $\Var{X_i}$, which is
$O(\card{V_i})$.

While efficient, \GreedyNaive\ tends to produce poor solutions in
practice (see Section~\ref{sec:expr}). Intuitively, \GreedyNaive\
assumes that cleaning the value with the highest variance reaps the
most benefit. At first glance, this estimation makes sense for both
\MinVar\ and \MaxPr: the most uncertain object value may contribute
the most to the query function result uncertainty, and a random draw
of this value may cause the largest deviation from the original query
function result. However, both assumptions easily fail in practice.
Recognizing \GreedyNaive's shortcoming of ignoring the objective when
deciding what to clean, we next show a different greedy method that
does consider the objective in its decision.

\mparagraph{Estimating benefits from optimization objectives} A better
strategy is to derive \Greedy's benefit estimation function from the
actual optimization objectives. We call the resulting greedy
algorithms \Greedy\MinVar\ and \Greedy\MaxPr. Recall that $T$ denotes
the set of objects chosen so far. For $\obj_i \in \Objs \setminus T$,
let $\delta_i$ denote the change in the objective if $T$ changes to
$T\cup\{\obj_i\}$. Let $\beta(\obj_i) = -\delta_i$ for \Greedy\MinVar\
and $\beta(\obj_i) = \delta_i$ for \Greedy\MaxPr.

Next, we show an example that
illustrates why \Greedy\MinVar\ performs better than \GreedyNaive\,
even in the simple case where $X_i$'s are mutually independent with unit cleaning cost,
and the query function is symmetric in them.
\begin{example}[\GreedyNaive\ vs \Greedy\MinVar]
  \label{eg:greedy-suboptimal}
  Recall the database setup in Example~\ref{eg:diff-objective}, where
  $X_1$ and $X_2$ are independently and uniformly distributed over
  $\{0, {1\over2}, 1, {3\over2}, 2\}$ and
  $\{{1\over3}, 1, {5\over3}\}$, resp. Suppose objects have unit
  cleaning cost and we have budget to clean only one object.
  \GreedyNaive\ will choose to clean $X_1$ because
  $\Var{X_1}>\Var{X_2}$.

  Consider \MinVar\ with query function $\One{X_1+X_2<{11\over12}}$.
  Note that this function returns $1$ for only two realizations of
  $(X_1,X_2)$, namely $(0,{1\over3})$ and $({1\over2},{1\over3})$.
  Thus,
  $\Var{\One{X_1+X_2<{11\over12}}} =
  {2\over5}\cdot{1\over3}\cdot(1-{2\over5}\cdot{1\over3}) =
  {26\over225}$.

  \Greedy\MinVar\ decides the item to clean by computing, for each item,
  the fraction of the variance improvement over the cost.

  If we clean $X_1$:
  \begin{itemize}
  \item With probability $2\over5$, $X_1\in\{0,{1\over2}\}$, so
    $\One{X_1+X_2<{11\over12}} = 1$ with probability
    $\Prob{X_2={1\over3}} = {1\over3}$. Therefore,
    $\Var{\!\!\One{\!X_1\!+\!X_2\!<\!{11\over12}}\!\!\mid\!\! X_1\!\!\in\!\!\{0,{1\over2}\}}\!\!=\!\!
    {1\over3}(1-{1\over3})\!\!=\!\!{2\over9}$.
  \item With probability $3\over5$, $X_1\ge1$, so
    $\One{X_1+X_2<{11\over12}} = 0$ for certain.\\Therefore,
    $\Var{\!\!\One{\!X_1\!\!+\!\!X_2\!<\!{11\over12}}\!\mid\!X_1\!\ge\!1}\!\!=\!\!0$.
  \end{itemize}
  Overall, the expected variance after cleaning $X_1$ is
  ${2\over5}\cdot{2\over9}={4\over45}$, and the improvement
  is ${26\over225}-{4\over45}\approx 0.0266$

  On the other hand, if we clean $X_2$ instead:
  \begin{itemize}
  \item With probability $1\over3$, $X_2={1\over3}$, so
    $\One{X_1+X_2<{11\over12}} = 1$ with probability
    $\Prob{X_1\in\{0,{1\over2}\}} = {2\over5}$. Therefore,
    $\Var{\One{X_1+X_2<{11\over12}} \mid X_2={1\over3}} =
    {2\over5}(1-{2\over5}) = {6\over25}$.
  \item With probability $2\over3$, $X_2\ge1$, so
    $\One{X_1+X_2<{11\over12}} = 0$ for certain.\\Therefore,
    $\Var{\One{\!X_1\!+\!X_2\!<\!{11\over12}}\!\mid\! X_2\!\ge\!1\!} = 0$.
  \end{itemize}
  Overall, the expected variance after cleaning $X_2$ is
  ${1\over3}\cdot{6\over25}={2\over25}$, and the improvement
  is ${26\over225}-{2\over25}=0.0355$.

  Hence, \Greedy\MinVar\ chooses to clean $X_2$.
  The expected uncertainty after cleaning $X_2$ (${2\over25}$)
  is lower than that of cleaning $X_1$ ($4\over45$).
  In other words, \Greedy\MinVar's choice of cleaning $X_2$ is better
  than the choice of \GreedyNaive's choice.
\end{example}

Since $\beta(\cdot)$ depends on $T$, we need to evaluate
$\beta(\cdot)$ in every iteration. Therefore, a straightforward
implementation of \Greedy\MinVar\ and \Greedy\MaxPr\ would have a time
complexity of $O(n^2\gamma)$, where $\gamma$ is the complexity of
computing the objective function. This complexity is highly dependent
on the forms of the data distribution and query function. Without any
assumption about such forms, a brute-force implementation would
enumerate all possible realizations of $\*X$, implying that
$\gamma = O(\card{\*V})$, which is exponential in $n$. To avoid this
high complexity, one possibility is to estimate $\delta_i$ using Monte
Carlo methods. Another possibility is to use more efficient
algorithms for certain forms of data distributions and query
functions; we defer that discussion to the next subsection.

\subsection{Modular Objectives}
\label{sec:algos:modular}

We now examine some practical cases when we can prove good
  theoretical bounds for the greedy algorithms, or devise algorithms
  with even better guarantees.  We start with a simple case where the
optimization objective is essentially linear.  An objective for
\MinVar\ or \MaxPr\ is \emph{modularizable over $\Objs$} if it is
equivalent to maximizing $\sum_{\obj_i \in T} w_i$ for some
$\*w = (w_1, w_2, \ldots, w_n) \in \Reals^n$ not dependent on $T$.
Modular objectives have a number of applications in practise.  For
example, window aggregate comparison claims can be expressed as linear
claim queries.  In this case, notice that the fairness of a claim is a
linear function, in the form $\*a\*X$ where
$\*a = (a_1, a_2, \ldots, a_n)$.  Hence, the problem of minimizing the
variance of fairness of linear claims is a special case of optimizing
a modular function over a weighted constraint.

The next lemma shows some conditions where \MinVar\ and \MaxPr\
problems have modular objectives.  All formal proofs can be found in Appendix~\ref{appndx:modular}.

\begin{lemma}\label{lemma:linear-modular}
  If components of $\*X$ are pairwise uncorrelated and $f(\*X)$ is
  affine (i.e., $f(\*X) = b + \*a \*X$, where $\*a = (a_1, a_2, \ldots, a_n)$), then \MinVar\ has a modularizable
  optimization objective, with $w_i = a_i^2 \Var{X_i}$.

  If components of $\*X$ are independently and normally distributed
  and centered around their current values (i.e.,
  $X_i \sim N(u_i, \sigma_i^2)$), and $f(\*X)$ is affine (i.e.,
  $f(\*X) = b + \*a \*X$, where $\*a = (a_1, a_2, \ldots, a_n)$),
  then \MaxPr\ has a modularizable optimization objective, with
  $w_i = a_i^2
  \sigma_i^2$.  
\end{lemma}

It is easy to see that \MinVar\ with modularizable
objective is equivalent to the standard minimum knapsack problem: given a
cost budget $C'$, choose $T' \subseteq \Objs$ to minimize
$\sum_{\obj_i \in T'} w_i$ subject to $\sum_{\obj_i \in T'} c_i \geq C'$.
Similarly, the \MaxPr\ problem with modularizable
objective is equivalent to the standard maximum knapsack problem:
given a cost budget $C$, choose $T \subseteq \Objs$ to maximize
$\sum_{\obj_i \in T} w_i$ subject to $\sum_{\obj_i \in T} c_i \leq C$.
This observation shows that \MinVar\ and \MaxPr\ are $\NP$-hard problems.
Using \cite{bentz2016note} we can get the following results.


\begin{lemma}\label{corollary:modular-minvar}
  For \MinVar, suppose the components of $\*X$ are pairwise
  uncorrelated and $f(\*X)$ is affine. Let $t$ denote the complexity
  of computing each $\Var{X_i}$. Then:
  \begin{itemize}
  \item An optimal solution can be computed
    in pseudo-polynomial time, more precisely, $O(n(t+C))$.
  \item For a parameter $\epsilon>0$,
    an $(1+\epsilon)$-approximate solution can be computed in
    $O(nt+n^3/\epsilon)$ time.
  \end{itemize}
\end{lemma}

Next we show the results for the \MaxPr\ problem.
We can take an exact pseudo-polynomial time algorithm for the maximum Knapsack problem
and attain an exact solution for the \MaxPr\ problem.
Furthermore, an approximate solution for the equivalent Knapsack problem can be used to
find an approximation solution to our original \MaxPr\ problem.
\begin{lemma}\label{corollary:modular-maxpr}
  For \MaxPr, suppose the components of $\*X$ are independently and
  normally distributed and centered around their current values, and
  $f(\*X)$ is affine. Then:
  \begin{itemize}
  \item An optimal solution can be computed
    in pseudo-polynomial time, or, more precisely, $O(nC)$.
  \item Let $\OPT$ denote the optimal objective function value.
  There is an algorithm that runs in $O(\frac{n^3}{\epsilon})$ time
  and return a value $A$ such that $A=O(\OPT)$, if $\OPT>0.05$.
  \end{itemize}
\end{lemma}

\mparagraph{Greedy for modularizable objectives}
By Lemma~\ref{lemma:linear-modular}, the benefit estimation function
for \Greedy\MinVar\ and \Greedy\MaxPr\, for the cases they cover, is
simply $\beta(\obj_i) = a_i^2 \Var{X_i}$, the variance of $X_i$
weighted by (the square of) its contribution to the query
function---this estimation is in fact exact. Since the benefit does
not depend on the objects already chosen, we can sort $\Objs$ upfront
by $\beta(\obj_i)/c_i$, without computing $\argmax$ in each iteration
of the loop. Therefore, the running times of \Greedy\MinVar\ and
\Greedy\MaxPr\ for these cases are the same as that of \GreedyNaive,
which is $O(n(t+\log n))$, where $t = O(\card{V_i})$ for
\Greedy\MinVar\ and $t = O(1)$ for \Greedy\MaxPr.

A well-known result on the knapsack problem is that \Greedy\ achieves
$2$-approximation. Thus, by Lemmas~\ref{lemma:linear-modular},
\Greedy\MinVar\ provides a $2$-approximation.
Furthermore, by doing a simple modification in the proof of Lemma~\ref{corollary:modular-maxpr}
we can show that \Greedy\MaxPr\ provides a constant approximation for the \MaxPr\ problem
(similarly to the second part of Lemma~\ref{corollary:modular-maxpr}).

\subsection{General Query Functions}
\label{sec:algos:submodular}
Now we consider the general case with arbitrary query
  function $f$.  Interestingly, it turns out that as long as the
  $X_i$'s are mutually independent, \MinVar's objective function is
  \emph{submodular} regardless of what the query function $f$ is. A
set function $g$ is submodular if for any set $A\subset B$, and
element $x\notin B$, it holds that
$g(A\cup x)-g(A)\geq g(B\cup x)-g(B)$.  This powerful observation
enables us to draw techniques from the rich literature on submodular
function optimization \cite{fujishige2005submodular,
  iwata2001combinatorial, nemhauser1978analysis,
  svitkina2011submodular}.  These observations would lead to
approximation algorithms for minimizing the expected variance of
uniqueness, robustness or any other function over the claims. Notice
that the results hold for any claim query (not only linear claim
queries), as long as the random distributions $X_i$'s are mutually
independent.

Our main technical contributions are the following: If the random
distributions $X_i$'s are mutually independent then i)~we show that
the objective of \MinVar\ is non-increasing and submodular regardless
of what the query function is, and ii)~we map \MinVar\ to a
minimization problem with a non-decreasing submodular objective
function and a linear lower bound cost constraint.  Finally, (ii)
allows us to use the algorithms in~\cite{iyer2013submodular} and get
efficient approximation algorithms for \MinVar.

In the following, let
$$\EVar{T}\!\!=\!\!\sum_{\*v \in \*V_T}\!\Prob{\*X_T\!\!=\!\!\*v\!} \cdot\Var{\!f(\*X)
  \!\!\mid\!\! \*X_T \!\!=\!\*v}$$
denote the objective function of \MinVar.

We first start with another observation that $\EVar{\cdot}$ is
non-increasing, which holds in general, regardless of data
distribution and query function.  Then, we show that $\EVar{\cdot}$ is
submodular as long as the $X_i$'s are mutually independent.  The
proofs of the next lemmas can be found in Appendix~\ref{appndx:proofsSub}.
\begin{lemma}\label{lemma:non-increasing}
  The objective function of \MinVar\ is monotone non-increasing in
  $T$; i.e., for all $T \subseteq \Objs$ and $\obj' \in \Objs$,
  $\EVar{T} \ge \EVar{T \cup \{\obj'\}}$.
\end{lemma}
\begin{lemma}\label{lemma:submodular}
  If components of $\*X$ are mutually independent, then the objective
  function of \MinVar\ is submodular in $T$; i.e., for all
  $T \subset T' \subset \Objs$ and $\obj_j \in \Objs \setminus T'$,
  $\EVar{T \cup \{\obj_j\}} - \EVar{T} \ge \EVar{T' \cup \{\obj_j\}} -
  \EVar{T'}$.
\end{lemma}

Therefore, \MinVar\ with mutually independent $X_i$'s is a problem of
minimizing a non-increasing submodular function with a linear cost
upper bound constraint. Next, we show that the \MinVar\ problem can
be mapped to a problem of minimizing a non-decreasing submodular
function with a linear cost lower bound constraint.  That will allow
us to use known algorithms from the literature of submodular
optimization.  The key idea is that instead of choosing the subset $T$
of objects to clean, we choose the subset $\overline{T}$ of objects to
\emph{not} clean (the cost constraint is complemented accordingly).
Let \MinVarComplement\ be the problem defined as follows: Choose
$\overline{T} \subseteq \Objs$ to minimize
$\EVarComplement{\overline{T}} =\EVar{\Objs \setminus \overline{T}}$
subject to $\sum_{\obj_i \in \overline{T}} c_i \ge \overline{C}$,
where $\overline{C} = \smash{\left(\sum_{\obj_i \in \Objs} c_i\right)- C}$.
The proof of the next lemma can be found in Appendix~\ref{appndx:proofsSub}.

\begin{lemma}\label{lemma:Compl}
  The \MinVar\ problem can be mapped to \MinVarComplement, with
  non-decreasing and submodular $\EVarComplement{\cdot}$.
\end{lemma}

Iyer and Bilmes in~\cite{iyer2013submodular} propose efficient approximation algorithms for the problem of minimizing a non-decreasing submodular function with a
submodular lower bound constraint. Notice that the \MinVarComplement\ problem has a linear lower bound constraint which is a (sub)modular function.
Hence, we can use the algorithms in~\cite{iyer2013submodular} to solve the \MinVarComplement\ problem and hence the \MinVar\ problem (from Lemma~\ref{lemma:Compl}).
In particular, they present an algorithm with approximation ratio
$O(\frac{H}{1-\kappa})$, where $\kappa=1-\min_{o_i\in \Objs}
\frac{\EVar{\emptyset}-\EVar{\{o_i\}}}{\EVar{O\setminus\{o_i\}}}$ (curvature of function $\EVar{\cdot}$), and $H$, in our case, is the approximation
ratio of an algorithm that minimizes a modular objective with a linear lower bound constraint (Knapsack problem).
Using~\cite{bentz2016note} we can get in polynomial time a $O(1)$-approximation for the minimization knapsack problem, so that gives a $O(\frac{1}{1-\kappa})$-approximation for the \MinVar\ problem.
In case that $\kappa=1$, we can use the observations in~\cite{iyer2013submodular} to get a $O(\sqrt{n}\log n \sqrt{H})=O(\sqrt{n}\log n)$-approximation algorithm for the \MinVar\ problem.

\remove{
We are interested in two particular algorithms presented in~\cite{iyer2013submodular}. The first one has approximation ratio $O(\sqrt{n}\log n)$,
and the second one $O(\frac{H}{1-\kappa})$, where $\kappa=1-\min_{o_i\in \Objs}
\frac{\EVar{\emptyset}-\EVar{\{o_i\}}}{\EVar{O\setminus\{o_i\}}}$ (curvature if function $\EVar{\cdot}$), and $H$, in our case, is the approximation
ratio of an algorithm that minimizes a modular objective with a linear lower bound constraint (Knapsack problem).
Using~\cite{bentz2016note} we can get in polynomial time a $O(1)$-approximation for the minimization knapsack problem, the second algorithm from~\cite{iyer2013submodular} gives a $O(\frac{1}{1-\kappa})$-approximation for the \MinVar\ problem.
}

The running time of the above algorithms is polynomial assuming that
$\EVar{\cdot}$ can be computed in polynomial time.
However, in the worst case $\EVar{\cdot}$ cannot be computed efficiently.
There are instances of the $\EVar{\cdot}$ function that are $\# P$-hard to compute exactly
and no approximation algorithm is known \cite{kleinberg2000allocating}.
In the next section we show how the function $\EVar{\cdot}$ can be computed efficiently
in practical applications of fact checking.

Putting everything together we obtain the following result.
\begin{theorem}
\label{theorem:submodular}
  \MinVar\ has an
  $O({1\over1-\kappa})$-approximation algorithm that runs in
  polynomial time, assuming that the
  $\EVar{\cdot}$ function can be computed in polynomial time,
  where $\kappa$ is the curvature of $\EVar{\cdot}$. If $\kappa=1$, the approximation ratio is $O(\sqrt{n}\log n)$.
\end{theorem}
In the special case where objects have unit cleaning cost, one option
is to apply a technique described in~\cite{svitkina2011submodular,
  hayrapetyan2005unbalanced} and get a \emph{bi-criteria}
approximation algorithm.  In particular we can get a set $T$ such that
$\EVar{T}\leq \frac{1}{\alpha}\EVar{T^*}$ and
$\sum_{\obj_i\in T}c_i\leq \frac{1}{1-\alpha}C$, for any $0<\alpha<1$.

\subsection{Application in Fact Checking}
\label{sec:algos:check}

We now turn back to fact-checking and show how specific problems in
this application domain can be solved using the algorithms proposed
earlier in this section.  We then return to the comparison between
ascertaining claim quality and finding counters, and show when these
two goals may align with each other.

\mparagraph{General claims}
As shown in Section~\ref{sec:model:check}, we formulate the problems
of cleaning data to ascertain claim quality and to find counters as
\MinVar\ and \MaxPr, respectively. Without any assumption on the data
distribution or type of the claims, we can only solve the general
instances of \MinVar\ and \MaxPr\ using \Greedy\MinVar\ and
\Greedy\MaxPr, but without any theoretical guarantee on optimality.
However, if we assume independent $X_i$'s---in other words, errors in
data values are independent---then by
Theorem~\ref{theorem:submodular}, we can apply the techniques from
Iyer and Bilmes~\cite{iyer2013submodular} to \MinVar\ to obtain an
efficient algorithm with approximation guarantees.
In particular, for any claim, we show that we can compute
  the $\EVar{}$ function efficiently for fairness, uniqueness, and
  robustness, the three measures of claim quality introduced in
  Section~\ref{sec:model} (and in~\cite{wu2014toward}).  This result
  implies that we can indeed run the algorithm in
  Theorem~\ref{theorem:submodular} for \MinVar\ in polynomial time for
  fact-checking. We prove the next theorem in Appendix~\ref{appndx:ApplFactChecking}.

\begin{theorem}\label{th:runtimeDepClaims}
  Let $V$ be the maximum support of distributions in $\*X$,
    and $W$ be the maximum number of objects referenced by each
    claim. Assuming that the components of $\*X$ are independent and $q(\*u)$ for each claim $q\in Q$ can be
    computed in $O(W)$ time, then for any set $T\subseteq \Objs$, the
    $\EVar{T}$ of $\bias{q^\circ(\*u)}{\*X}$,
    $\duplicity{q^\circ(\*u)}{\*X}$, and
    $\fragility{q^\circ(\*u)}{\*X}$ can be computed in
    $O(m^2V^{3W}W+n)$ time.
\end{theorem}
Note that in practice $W$ is a small constant (e.g., $W=8$
  in Giuliani's claims), so we can compute the $\EVar{\cdot}$ function
  and eventually run the algorithm from
  Theorem~\ref{theorem:submodular} in $O(\polyn(n,m,V))$.  Also note
  that if we always clean the values referenced by the original claim
  $q^\circ$ upfront, then the time to compute $\EVar{\cdot}$ can be
  improved to $O(mLV^{2W}W)$, where $L$ is the maximum degree of a
  claim (the degree of claim $q$ is defined as the number of claims
  that share at least one object with $q$).

\mparagraph{Linear claims}
If we further consider a common class of claim functions that are
linear, more efficient algorithms become available.  \emph{Linear
  claim functions} are those that can be expressed in the form
$\*a\*X$ where $\*a = (a_1, a_2, \ldots, a_n)$ is a vector of weights
associated with each object value. For example, the window aggregate
comparison claims considered in Example~\ref{eg:window-aggr-compare}
are linear: $a_i = -1$ if $\obj_i$ belongs to the first window but not
the second, $1$ if $\obj_i$ belongs to the second window but not the
first, and $0$ otherwise. In general, any SQL aggregation query over
selections and joins is linear, provided that selection and join
conditions involve only attribute values that are certain and
therefore not included in $\*X$.

Suppose the original claim function $q^\circ$ is linear; let
$\*a^\circ$ denotes the weights used by $q^\circ$. Perturbations
$q_1, \ldots, q_m$ of a linear query are linear too; let $\*A$ be an
$m \times n$ matrix where row $\*{a}_{k,*}$ denotes the weights used
by $q_k$; i.e.,
$q_k(\*X) = a_{k,1} X_1 + a_{k,2} X_2 + \cdots + a_{k,n} X_n$. Let
$\*a_{*,i}$ denote the $i$-th column of $\*A$, i.e., the weights used
for $X_i$ by the $m$ perturbation queries. Further suppose that
relative strength function $\Delta(\cdot,\cdot)$ simply subtracts its
inputs, which is a natural choice for linear claim functions.

If we want to ascertain the original claim's fairness, we need to
solve \MinVar\ with query function $\bias{q^\circ(\*u)}{\*X}$
(Section~\ref{sec:model:check}). We note that this query function is
linear given linear claim functions. More specifically,
$\bias{q^\circ(\*u)}{\*X} = \*w\*X$ where
$w_i = \sum_{1 \le k \le m} s(q_k)(a_{k,i} - a^\circ_i)$. Hence, as
long as the components of $\*X$ are pairwise uncorrelated, the query
function is modular and can be solved efficiently as a knapsack
problem (Section~\ref{sec:algos:modular}).

Note that for the task of ascertaining claim qualities, linear claim
functions do not always imply linear query functions for \MinVar. For
uniqueness and robustness, for example, the query functions
$\duplicity{q^\circ(\*u)}{\*X}$, $\fragility{q^\circ(\*u)}{\*X}$
introduce non-linearity through with their additional use of indicator
and quadratic functions, so for \MinVar\ with these two query
functions, the results for modular objective functions do not apply.
Instead, we can use Theorems~\ref{theorem:submodular}
  and~\ref{th:runtimeDepClaims} for constant $W$ and find a good
  approximation for \MinVar\ with $\duplicity{q^\circ(\*u)}{\*X}$ or
  $\fragility{q^\circ(\*u)}{\*X}$ in polynomial time.

Now for the task of finding counters, we need to solve \MaxPr\ with
query function $\bias{q^\circ(\*u)}{\*X}$
(Section~\ref{sec:model:check}).  As discussed earlier, this query
function is linear given linear claim functions. Therefore, we can
solve this problem as a knapsack problem if the components of $\*X$
are independently and normally distributed and centered around their
current values (Section~\ref{sec:algos:modular}).

\mparagraph{Ascertaining claim quality vs.\ finding counters}
Finally, we return to the comparison between ascertaining
  claim quality and finding counters.  As seen in
  Example~\ref{eg:diff-objective} of Section~\ref{sec:model:check},
  the two objectives in general differ.  However, we show here that
  they turn out to agree with each other for linear claim queries.
  Following the analysis earlier in this subsection on linear claim
  queries, it is not hard to see that \MinVar\ with query function
  $\bias{q^\circ(\*u)}{\*X}$ (i.e., the goal of ascertaining claim
  fairness) and \MaxPr\ with query function $\bias{q^\circ(\*u)}{\*X}$
  (i.e., the goal of finding counters) are aligned for linear claim
  functions when the components of $\*X$ are independently and
normally distributed and centered around their current values.  In
fact, we can extend this observation and show a stronger result for
any multivariate normal distribution, without making the
  independence assumption.  The full proof can be found
in Appendix~\ref{appndx:ApplFactChecking}.
\begin{theorem}\label{theorem:minvar-maxpr-equivalence}
  If $\*X$ follows a multivariate normal distribution centered around
  the current values $\*u$, and if all claim functions
  $q^\circ, q_1, \ldots, q_m$ are linear and the relative strength
  function $\Delta(\cdot,\cdot)$ is the subtraction operation, then
  for query function $\bias{q^\circ(\*u)}{\*X}$, the optimal solutions
  to \MinVar\ and \MaxPr\ are the same.
\end{theorem}


\section{Experiments}
\label{sec:expr}

In this section, we experimentally evaluate the effectiveness of our
proposed algorithms in achieving their objectives.  We also
demonstrate their efficiency, and examine how different
optimization objectives can lead to different outcomes, with
implications to the practice of fact-checking.
Our experiments use realistic claim functions and a combination of
synthetic and real data.  We use synthetic value uncertainty and cost
distributions when such information is unavailable, and when we want
to evaluate with different distributions.  Details on the datasets are
presented below; claim functions will be specified for individual
experiments later.

\emph{\textbf{Adoptions}} is a dataset derived from the number of
adoptions in the New York City during 1989--2014.  While the numbers
were real, there was no published error model, so we assume that the
data contains error modeled as follows.  $X_i$, the number of
adoptions in a particular year, follows a normal distribution with
mean $u_i$, the current (reported) value, and standard deviation drawn
uniformly from $[1,50]$.  We assume that the cost of cleaning each
$X_i$ is drawn uniformly at random from $[1,100]$.

A\emph{\textbf{CDC-firearms}} and \emph{\textbf{CDC-causes}}
  are real datasets complete with error models published by the
  Centers for Disease Control and Prevention (CDC)~\cite{CDCData}.
  CDC routinely collects statistics on injuries and deaths by various
  causes through established sampling methods, and publishes the data
  along with statistics like standard errors, coefficients of
  variation, and $95\%$ confidence intervals.  Note that sampling
  procedures used by CDC ensure that the errors are independent and
  follow approximately normal distributions.  For \emph{CDC-firearms},
  we get the estimated numbers of nonfatal firearms injuries in the
  USA during 2001--2017, along with the standard deviations.  For
  \emph{CDC-causes}, in addition to nonfatal firearms injuries, we
  also get the data on injuries due to transportation, as well as
  drownings and falls over the same time period; this results in a
  larger dataset with $68$ values and enables more varieties of claim
  queries.  We do not know the actual costs of cleaning, but a
  reasonable assumption is that acquiring older historical data is
  more expensive.  Therefore, we generate cleaning costs in a way such
  that they decrease with recency: the cost of cleaning a value from
  the year 2001 is a random number in $195$--$200$, the cost for 2002
  is in $190$--$195$, etc.

\emph{\textbf{Synthetic datasets}} $\mathsf{UR}_x$, $\mathsf{LN}_x$,
and $\mathsf{SM}_x$ are used to explore how different value
distributions and dataset sizes affect various algorithms.  For each
value $X_i$, we first choose the size of its support uniformly at
random from $[1,6]$.  Then, we generate the distribution for $X_i$
with one of the following methods:
\begin{itemize}
\item $\mathsf{UR}_x$ generates fairly ``random'' distributions.  We
  choose elements of $\mathrm{supp}(X_i)$ uniformly at random from
  $[1,100]$ without replacement.  For each element, we assign its
  probability in proportion to a number drawn uniformly at random from
  $(0,1]$ (normalized such that the total over all of
  $\mathrm{supp}(X_i)$ is $1$).
\item $\mathsf{LN}_x$ generates skewed but unimodal value
  distributions.  We start with a log-normal distribution with
  parameters $\mu=0$ and $\sigma$ chosen uniformly at random in
  $(0,1]$.  We quantilize distribution into as many equal-probability
  intervals as $\card{\mathrm{supp}(X_i)}$, and choose elements of
  $\mathrm{supp}(X_i)$ to be close to the right ends of these
  intervals.  For each element, we then assign its probability in
  proportion to its probability density in the log-normal distribution
  (again, normalized to sum to $1$).  Note that resulting range is
  typically much smaller than the other two methods (which are based
  on $[1,100]$).
\item $\mathsf{SM}_x$ generates more complex multimodal distributions.
  We choose elements of $\mathrm{supp}(X_i)$ in the same way as
  $\mathsf{UR}_x$, but for each element, we assign its probability in
  proportion to a random number in $(1,0.1] \cup [0.9,1]$, i.e.,
  either low or high.
\end{itemize}
For cleaning cost, we draw it uniformly at random from $[1, 10]$ for
each object.  We have also experimented with a different cost
distribution where individual costs are more extreme (either $1$ or
$10$).  It led to similar results and revealed no additional insights;
hence we omitted the results here.

\subsection{Effectiveness for Modular Objectives}
\label{sec:expr:modular}

We start by evaluating the effectiveness of our approach for the
(simpler) case of modular objectives discussed in
Section~\ref{sec:algos:modular}. We will focus on the scenario of
minimizing uncertainty in the fairness measure of a window aggregate
comparison claim (Example~\ref{eg:window-aggr-compare}); our algorithm
for this scenario is \Greedy\MinVar.
(We omit the results for \Greedy\MaxPr\ because they are similar.)

\mparagraph{Workloads}
We consider Giuliani's adoption claim
(Example~\ref{eg:window-aggr-compare}) over \emph{Adoptions}.  The
claim function subtracts the total adoption numbers over two four-year
periods back-to-back (1993--1996 vs.\ 1989--1992).  For analysis of
fairness, we consider $18$ perturbations of the original claim, all
having the same form but each ending with a different year.  To
capture the intuition that important perturbations are those closest
to the original in terms of the time periods compared, we let the
sensibility of a perturbation decay exponentially (at rate
$\lambda = 1.5$) over its distance to the original claim (as measured
by the number of years between the endpoints of their comparison
periods).  Overall, the query function corresponding to fairness is
complex albeit still linear, where the weight on each $X_i$
incorporates the sensibilities of perturbations involving $X_i$.

We also consider claims over \emph{CDC-firearms} and
  \emph{CDC-causes}.  For \emph{CDC-firearms}, the claim function
  subtracts the numbers of firearms injuries over two four-year
  periods back-to-back, e.g., 2001--2004 vs.\ 2005--2008; such claims
  can be used to argue, for example, that current policy is more (or
  less) effective than the previous one.  We consider $10$
  perturbations, with sensibility set up similarly as Giuliani's
  adoption claim.  For \emph{CDC-causes}, the claim function
  aggregates injury numbers across causes for comparison: ``the number
  of injuries due to transportation is more than 30\% of all other
  causes combined over the last 2-year period.''  We consider $16$
  perturbations, again with a similar sensibility setup.


\mparagraph{Algorithms compared}
%
\begin{itemize}
\item $\mathsf{Random}$ simply chooses a next object to clean
  uniformly at random (with no replacement), until the budget is
  exceeded.
\item $\GreedyNaive\mathsf{CostBlind}$ sorts the objects according to
  the variances in individual $X_i$'s, and chooses them in descending
  order of uncertainty to clean, until the budget is exceeded.
\item \GreedyNaive\ (discussed in Section~\ref{sec:algos:greedy})
  estimates the benefit of cleaning $X_i$ simply as its variance, and
  cleans objects in descending order of benefit/cost ratios.
\item \Greedy\MinVar\ (our proposed algorithm) estimates the benefit
  of each $X_i$ from the optimization objective---which considers the
  claim function as well as the structure and sensibilities of
  perturbations---and cleans objects in descending order of
  benefit/cost ratios.
\item $\mathsf{Optimum}$ (discussed in
  Lemma~\ref{corollary:modular-minvar}) solves the same knapsack
  problem as \Greedy\MinVar, but uses dynamic programming to find the
  optimum solution in pseudo-polynomial time, which is much slower
  than \Greedy\MinVar.
\end{itemize}

\begin{figure}[h]
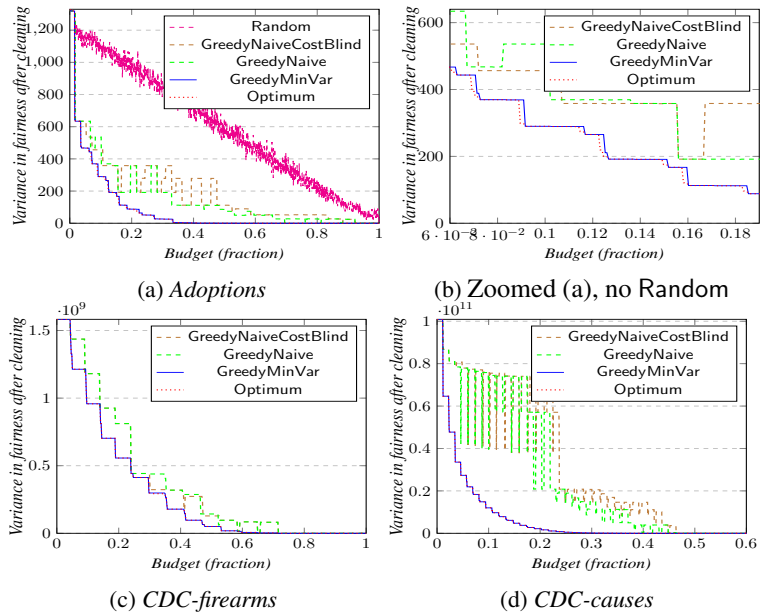

\centering
  \subfloat[\emph{Adoptions}]{\input{figs/expr/perc/adoption-fairness2}}
  \subfloat[{\small Zoomed (a), no $\mathsf{Random}$}]{\input{figs/expr/perc/adoption-fairness-zoom2}}\\[-2.5ex]
  \subfloat[\emph{CDC-firearms}]{\input{figs/expr/perc/CDC_fairnessCorCosts}}
  \subfloat[\emph{CDC-causes}]{\input{figs/expr/perc/CDC_fairnessAllCorCosts_2}}
  \caption{\label{fig:expr:adoption-fairness}Effectiveness of
    algorithms in reducing uncertainty in claim fairness.}
\end{figure}

\mparagraph{Results and discussion}
Figure~\ref{fig:expr:adoption-fairness} compares the effectiveness of
the above algorithms in reducing uncertainty in claim fairness.  The
horizontal axis shows the cost budget as a percentage of the total
cost of cleaning all data.  Given a budget, we run each algorithm and
plot the uncertainty (variance) that remains in the fairness of the
original claim after cleaning the values chosen by the
algorithm.\footnote{\small Since $\mathsf{Random}$ randomly chooses
  data to clean, its result can vary significantly depending on its
  random choices.  Hence, we conduct $100$ runs of $\mathsf{Random}$
  and plot the average remaining variance; to keep the plot readable,
  we omit error bars (which are quite large).  For other algorithms,
  the remaining uncertainty is computed exactly given their
  (deterministic) choices.}  Note that the range of variance
  values depends on the dataset and claim in question, so we should
  focus on the relative performance of algorithms instead of absolute
  variance values.

From Figure~\ref{fig:expr:adoption-fairness}a, we see a large gap
between $\mathsf{Random}$ and other algorithms.  With
$\mathsf{Random}$, uncertainty decreases linearly with increasing
budget.  Other algorithms are able to reduce far more uncertainty even
with relatively low budgets; however, as budget increases, the
reduction in uncertainty eventually slows down as expected.  To better
show the differences among other algorithms, we omit $\mathsf{Random}$
from other figures; Figure~\ref{fig:expr:adoption-fairness}b is a
zoomed version of Figure~\ref{fig:expr:adoption-fairness}a.
On all datasets, we see that our algorithm, \Greedy\MinVar,
  is very effective and almost indistinguishable from
  $\mathsf{Optimum}$.  For example, in Giuliani's adoption claim,
with only $3\%$ of the total cleaning cost, it is able to reduce
uncertainty by a factor of $2.8$.  We also observe that
\Greedy\MinVar\ clearly outperforms its less sophisticated cousins
\GreedyNaive\ (which ignores the query function of interest) and
$\GreedyNaive\mathsf{CostBlind}$ (which further ignores differences in
cleaning costs) in all datasets.


\begin{figure}[h]
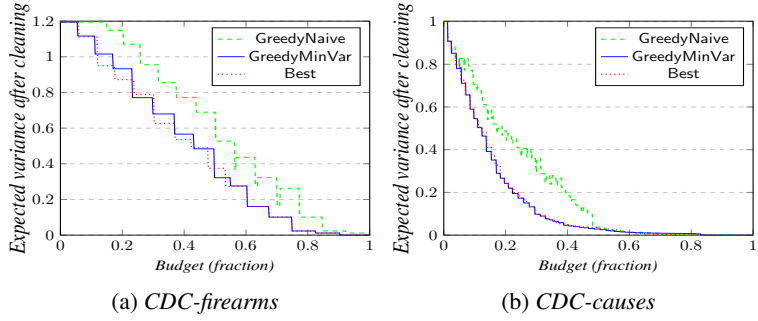

\centering
  \subfloat[\emph{CDC-firearms}]{\input{figs/expr/perc/CDC_UniquenessCorCosts}}
  \subfloat[\emph{CDC-causes}]{\input{figs/expr/perc/CDC_UniquenessAllCorCosts2}}
  \caption{\label{fig:expr:un:real}Effectiveness of algorithms
    in reducing uncertainty in claim uniqueness (CDC datasets).}
\end{figure}

\begin{figure}[h]
\centering
  \subfloat[$\Gamma=50$]{\input{figs/expr/perc/univ-unic-r50}}
  \subfloat[$\Gamma=100$]{\input{figs/expr/perc/univ-unic-r100}}\\[-3ex]
  \subfloat[$\Gamma=150$]{\input{figs/expr/perc/univ-unic-r150}}
  \subfloat[$\Gamma=200$]{\input{figs/expr/perc/univ-unic-r200}}\\[-3ex]
  \subfloat[$\Gamma=250$]{\input{figs/expr/perc/univ-unic-r250}}
  \subfloat[$\Gamma=300$]{\input{figs/expr/perc/univ-unic-r300}}
  \caption{\label{fig:expr:uniqueness:univ-unic}Effectiveness of
    various algorithms in reducing the uncertainty in the uniqueness
    of a claim asserting an aggregate result to be as small as
    $\Gamma$.  Value distributions are generated using
    $\mathsf{UR}_x$.}
\end{figure}

\begin{figure}[h]
\centering
  \subfloat[$\Gamma=3.0$]{\input{figs/expr/perc/logv-unic-r30}}
  \subfloat[$\Gamma=3.5$]{\input{figs/expr/perc/logv-unic-r35}}\\[-3ex]
  \subfloat[$\Gamma=4.0$]{\input{figs/expr/perc/logv-unic-r40}}
  \subfloat[$\Gamma=4.5$]{\input{figs/expr/perc/logv-unic-r45}}\\[-3ex]
  \subfloat[$\Gamma=5.0$]{\input{figs/expr/perc/logv-unic-r50}}
  \subfloat[$\Gamma=5.5$]{\input{figs/expr/perc/logv-unic-r55}}
  \caption{\label{fig:expr:uniqueness:logv-unic}Effectiveness of
    various algorithms in reducing the uncertainty in the uniqueness
    of a claim asserting an aggregate result to be as small as
    $\Gamma$.  Value distributions are generated using
    $\mathsf{LN}_x$.}
\end{figure}

\begin{figure}[h]
\centering
  \subfloat[$\Gamma=50$]{\input{figs/expr/perc/nonv-unic-r50}}
  \subfloat[$\Gamma=100$]{\input{figs/expr/perc/nonv-unic-r100}}\\[-3ex]
  \subfloat[$\Gamma=150$]{\input{figs/expr/perc/nonv-unic-r150}}
  \subfloat[$\Gamma=200$]{\input{figs/expr/perc/nonv-unic-r200}}\\[-3ex]
  \subfloat[$\Gamma=250$]{\input{figs/expr/perc/nonv-unic-r250}}
  \subfloat[$\Gamma=300$]{\input{figs/expr/perc/nonv-unic-r300}}
  \caption{\label{fig:expr:uniqueness:nonv-unic}Effectiveness of
    various algorithms in reducing the uncertainty in the uniqueness
    of a claim asserting an aggregate result to be as small as
    $\Gamma$.  Value distributions are generated using
    $\mathsf{SM}_x$.}
\end{figure}

\begin{figure}[h]
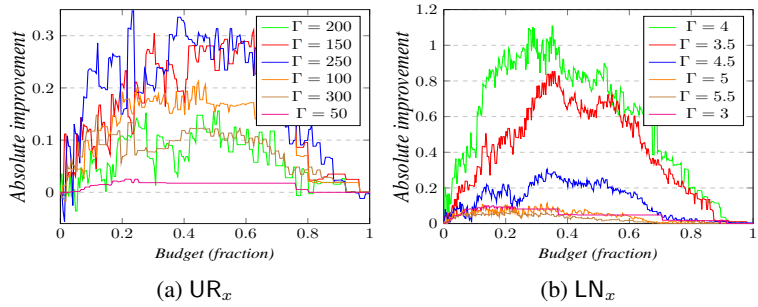

\centering
  \subfloat[$\mathsf{UR}_x$]{\input{figs/expr/perc/univ-unic-abs-improvement}}
  \subfloat[$\mathsf{LN}_x$]{\input{figs/expr/perc/logv-unic-abs-improvement}}
  \caption{\label{fig:expr:uniqueness:abs-improvement}Absolute
    improvement of \Greedy\MinVar\ over \GreedyNaive\ in reducing
    uncertainty, for the same scenarios in
    Figures~\ref{fig:expr:uniqueness:univ-unic}
    and~\ref{fig:expr:uniqueness:logv-unic}, respectively.}
\end{figure}

\subsection{Effectiveness for Non-Modular Objectives}
\label{sec:expr:submodular}

Here we focus on the scenario of minimizing uncertainty in claim
uniqueness and robustness, where the optimization objective
is no longer modular but the results of
Section~\ref{sec:algos:submodular} apply.

\mparagraph{Workloads}
For minimizing the uncertainty on uniqueness, for
  \emph{CDC-firearms} and \emph{CDC-causes} datasets, we consider a
  claim of the form: ``in the last two years, the number of injuries
  by firearms (or across four categories, resp.) is as low as
  $\Gamma$.''  To assess claim quality, we consider $8$ perturbations,
  each summing $2$ (or $8$, resp.) object values.  Intuitively,
  checking uniqueness involves counting how many perturbations yield
  results no higher than $\Gamma$.  For minimizing the uncertainty on
  robustness, for \emph{CDC-firearms} and \emph{CDC-causes} datasets,
  we consider a claim of the form: ``in the last two years, the number
  of injuries by firearms (or across four categories, resp.) is as
  high as $\Gamma'$.''  We consider the same number of perturbations
  as in the uniqueness case.  Intuitively, checking robustness
  involves assessing how easy it is to find a perturbation that yields
  a result much lower than $\Gamma'$.  Note that in all cases the
  query functions are non-linear.  The goal is to choose a set of
  values to clean to minimize the variance in uniqueness or
  robustness.

We also use the three synthetic datasets in order to study how various
value distributions affect the effectiveness of
algorithms.
The claim sums up $4$ consecutive object value and states that the sum
is as low as $\Gamma$ (for uniqueness) or as high as $\Gamma'$ (for
robustness).  For experiments on uniqueness, we generate small
datasets with $40$ uncertain values each, which make results easier to
interpret (Section~\ref{sec:expr:efficiency} experiments with larger
datasets to evaluate efficiency); $10$ perturbations of the original
claim are used to assess uniqueness. For experiments on
  robustness, we generate bigger datasets with $100$ uncertain values
  each; $25$ perturbations are used to assess robustness.
Note that the value of $\Gamma$ or $\Gamma'$ appearing in
  the original claim can affect both the initial uncertainty and how
  much reduction cleaning each value would bring, because certain sums
  can be more likely than others depending on the value distribution.
  To study this effect, we also test claims with different
  $\Gamma$/$\Gamma'$ values.

\mparagraph{Algorithms compared}
Besides \GreedyNaive\ and \Greedy\MinVar, we also implemented
$\mathsf{Best}$, the $O(\frac{1}{1-\kappa})$-approximation algorithm
from~\cite{iyer2013submodular} with theoretical guarantees described
in Section~\ref{sec:algos}.  Note that $\mathsf{Best}$ does not
guarantee an optimum solution, but it has the best known theoretical
guarantee; hence, we use it as a yardstick for comparison.
Since these algorithms work with discrete distributions, for
  CDC datasets, we discretize each normal distribution with the given
  mean and standard deviation using $6$ and $4$ discrete values for \emph{CDC-firearms} and \emph{CDC-causes}, respectively.

We note that there are many existing techniques for data cleaning, however it seems unlikely to compare our algorithms with them.
Please look Section~\ref{sec:related} for a more detailed discussion.
A different approach is to clean the data upfront and then apply
perturbation analysis for fact-checking; however, a consequence of
cleaning data without any guidance by the goal of fact-checking is
that we may end up cleaning a lot of data that do not help with our
goal.  This is particularly problematic to fact-checkers because in
practice they are severely constrained by time and resources.  The
\GreedyNaive\ algorithm we test in this section is in
fact a representative instance of this approach, which tries to
clean data to reduce overall uncertainty, but without taking into
consideration of the goal of fact-checking.

\mparagraph{Results and discussion}
Figure~\ref{fig:expr:un:real} shows how various algorithms
  decrease uncertainty in claim uniqueness with increasing budget for
  \emph{CDC-firearms} and \emph{CDC-causes}.  The horizontal axes show
  the budget, while the vertical axes show the (computed) expected
  variance after cleaning the values chosen by each algorithm.  We
  observe that $\mathsf{Best}$ and \Greedy\MinVar\ have almost the
  same performance and they outperform \GreedyNaive\ in all cases.
  For example, in Figure~\ref{fig:expr:un:real}b, with $20\%$ of the
  maximum budget, $\mathsf{Best}$ and \Greedy\MinVar\ find strategies
  that lead to half of the variance of what \GreedyNaive\ is able to
  achieve.  (Note that even though the range of variance values seems
  small, it needs to be interpreted in context---as the range of
  $\duplicity{\cdot}{\cdot}$ is also small in this case, even a
  variance of $1$ implies significant uncertainty.)

Figures~\ref{fig:expr:uniqueness:univ-unic}, \ref{fig:expr:uniqueness:logv-unic}, \ref{fig:expr:uniqueness:nonv-unic}
compare how various algorithms reduce uncertainty in claim uniqueness when value
distributions are generated using $\mathsf{UR}_x$, $\mathsf{LN}_x$,
and $\mathsf{SM}_x$, respectively.  Sub-figures further show how the
workload parameter $\Gamma$ affects the results.
The horizontal axes
show the budget, while the vertical axes show the (computed) expected
variance after cleaning the values chosen by each algorithm.

Overall, we
find \Greedy\MinVar\ and $\mathsf{Best}$ effective across different
value distributions, budgets, and claim parameters ($\Gamma$).
Despite being simpler, \Greedy\MinVar\ is at least comparable to
$\mathsf{Best}$, and sometimes better. Generally speaking, they
outperform the less sophisticated \GreedyNaive, and the lead can be
substantial.  There is only one case where $\mathsf{Best}$ is slightly
beaten by \GreedyNaive\ (Figure~\ref{fig:expr:uniqueness:univ-unic}b with
$\Gamma=200$ and enough budget), but even there \Greedy\MinVar\
consistently beats \GreedyNaive.

A second observation is the effect of $\Gamma$, which can be seen
across the subfigures in Figures~\ref{fig:expr:uniqueness:univ-unic}, \ref{fig:expr:uniqueness:logv-unic}, \ref{fig:expr:uniqueness:nonv-unic}.
Generally, the initial
uncertainty (when cleaning budget is $0$, i.e., no data is cleaned
yet) is the highest if $\Gamma$ is in the midrange where the indicator
functions in the query function could easily go either way.  For
example, in Figure~\ref{fig:expr:uniqueness:univ-unic}d, we see that
the initial variance is more than $1.6$ when $\Gamma=200$, but only
about $0.45$ in Figure~\ref{fig:expr:uniqueness:univ-unic}b when $\Gamma=100$ and less than $0.40$
in Figure~\ref{fig:expr:uniqueness:univ-unic}f when $\Gamma=300$.
The same trend is visible in
Figure~\ref{fig:expr:uniqueness:nonv-unic} where initial uncertainty
peaks around the same $\Gamma$ value, because both $\mathsf{UR}_x$ and
$\mathsf{SM}_x$ draw values from $[1,100]$.  The peak $\Gamma$ is only
around $4$ in Figure~\ref{fig:expr:uniqueness:logv-unic}, because the
high-probability value range is much smaller under $\mathsf{LN}_x$;
here, the decrease in initial uncertainty is slower to the right of
peak $\Gamma$ than to the left because of the skew in the underlying
log-normal distribution.

As a related observation, when uncertainty is low to begin with, the
advantages of \Greedy\MinVar\ and $\mathsf{Best}$ over \GreedyNaive,
in relative terms, are more pronounced.  This observation is
encouraging because in practice, most claims we are interested in
involve $\Gamma$ values that are out of ordinary, which correspond to
regions where \Greedy\MinVar\ and $\mathsf{Best}$ significantly
outperform \GreedyNaive.

While it may appear from these figures that the differences among
algorithms are small when the initial uncertainty is high, we note
that if we instead examine the improvement in uncertainty by\\
\Greedy\MinVar\ over \GreedyNaive\ in absolute terms, we would in fact
see bigger improvements in cases with larger initial uncertainty.
Figure~\ref{fig:expr:uniqueness:abs-improvement} shows the absolute
improvement (in the amount of expected variance reduced) of
\Greedy\MinVar\ over \GreedyNaive, for the same scenarios in
Figures~\ref{fig:expr:uniqueness:univ-unic} ($\mathsf{UR}_x$)
and~\ref{fig:expr:uniqueness:logv-unic} ($\mathsf{LN}_x$); the
scenario of Figures~\ref{fig:expr:uniqueness:nonv-unic}
($\mathsf{SM}_x$) is similar.  Each curve shows, for a specific value
of $\Gamma$, the improvement as a function of the budget.  The legends
list the $\Gamma$ values in descending order of the corresponding
initial uncertainties.  We see that this ordering is fairly consistent
with the ordering of the curves; i.e., a higher initial uncertainty
translates to a bigger absolute improvement of \Greedy\MinVar\ over
\GreedyNaive.  For example, in
Figure~\ref{fig:expr:uniqueness:abs-improvement}b, improvement is the
biggest for $\Gamma=4.0$, which has the peak initial uncertainty;
improvement is also big for $\Gamma=3.5$ and $\Gamma=4.5$, whose
initial uncertainties are the next highest.  In contrast, the absolute
improvement for $\Gamma=3$ is small, even though the relative
improvement shown in Figure~\ref{fig:expr:uniqueness:logv-unic}a is
huge.
The results in Figure~\ref{fig:expr:uniqueness:abs-improvement}a
for $\mathsf{UR}_x$  are not as clear as for $\mathsf{LN}_x$. However
we can observe that for $\Gamma=50$ the absolute improvement is very small,
while the relative improvement was huge as shown in Figure~\ref{fig:expr:uniqueness:univ-unic}b.
Figure~\ref{fig:expr:uniqueness:abs-improvement} also allows us to see
the effect of the budget constraint on the improvement of
\Greedy\MinVar\ over \GreedyNaive.  When the budget becomes either
very tight or very generous, the difference between \Greedy\MinVar\
over \GreedyNaive\ becomes smaller.  This effect is consistent with
intuition: a tight budget means limited options, while a generous
budget means choices matter less since most uncertainty will be
removed anyway.

To sum up, \Greedy\MinVar\ consistently does the best job in removing
uncertainty in uniqueness across various value distributions, budgets,
and other workload settings in these experiments.  In some cases it
even beats our yardstick, $\mathsf{Best}$, but that should not be
surprising since $\mathsf{Best}$'s guaranteed approximation factor
depends on the curvature of the objective function (as discussed in
Theorem~\ref{theorem:submodular}), which means $\mathsf{Best}$ may not
be optimum in practice.

\begin{figure}[h]
\centering
  \subfloat[CDC-firearms]{\input{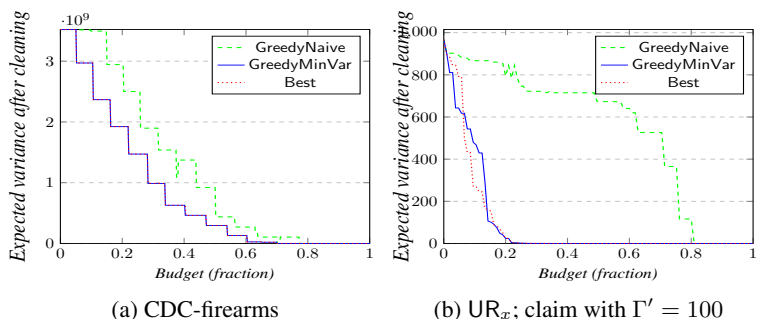}}
  \subfloat[$\mathsf{UR}_x$; claim with $\Gamma'=100$]{\begin{tikzpicture}[yscale=0.5, xscale=0.6]
\begin{axis}[
    y label style={at={(axis description cs:0.1,.5)},anchor=south},
    xlabel=\emph{Budget (fraction)},
    ylabel=\emph{{\Large Expected variance after cleaning}},
    xmin=0, xmax=1,
    ymin=0, ymax=1015,
    legend pos=north east,
    ymajorgrids=true,
    grid style=dashed,
]
\addplot[
    color=green, thick,
    dashed
    ]
    coordinates {
(0.000000, 967.124524) (0.009524, 902.500846) (0.019048, 902.499153) (0.028571, 902.497428) (0.038095, 902.496693) (0.047619, 892.660650) (0.057143, 882.268463) (0.066667, 882.268463) (0.076190, 878.455460) (0.085714, 867.158263) (0.095238, 867.147320) (0.104762, 867.138828) (0.114286, 867.139907) (0.123810, 867.131415) (0.133333, 867.131415) (0.142857, 867.131415) (0.152381, 867.131415) (0.161905, 860.256686) (0.171429, 860.256686) (0.180952, 860.256686) (0.190476, 854.219449) (0.200000, 794.429742) (0.209524, 844.201842) (0.219048, 784.412135) (0.228571, 844.143529) (0.238095, 784.353822) (0.247619, 746.214559) (0.257143, 746.214559) (0.266667, 722.382501) (0.276190, 722.249380) (0.285714, 721.925131) (0.295238, 721.925131) (0.304762, 721.792009) (0.314286, 721.921603) (0.323810, 721.921603) (0.333333, 715.004288) (0.342857, 721.921603) (0.352381, 715.004288) (0.361905, 715.004288) (0.371429, 715.004288) (0.380952, 715.004288) (0.390476, 714.845855) (0.400000, 715.004288) (0.409524, 714.845855) (0.419048, 714.871167) (0.428571, 714.845855) (0.438095, 714.871167) (0.447619, 714.845855) (0.457143, 714.871167) (0.466667, 714.871167) (0.476190, 714.871167) (0.485714, 714.871167) (0.495238, 674.616173) (0.504762, 674.616173) (0.514286, 673.145470) (0.523810, 673.145469) (0.533333, 672.651127) (0.542857, 672.651127) (0.552381, 672.625814) (0.561905, 672.625814) (0.571429, 672.625814) (0.580952, 640.513079) (0.590476, 640.513077) (0.600000, 640.513077) (0.609524, 620.120444) (0.619048, 620.068760) (0.628571, 526.679882) (0.638095, 526.628198) (0.647619, 526.628198) (0.657143, 526.628198) (0.666667, 526.493354) (0.676190, 526.493354) (0.685714, 526.481910) (0.695238, 526.481910) (0.704762, 526.481910) (0.714286, 365.429954) (0.723810, 365.429954) (0.733333, 365.429954) (0.742857, 365.360124) (0.752381, 365.360124) (0.761905, 116.421157) (0.771429, 116.421154) (0.780952, 116.421157) (0.790476, 116.421154) (0.800000, 115.907568) (0.809524, 0.521627) (0.819048, 0.008040) (0.828571, 0.008040) (0.838095, 0.007779) (0.847619, 0.007779) (0.857143, 0.007610) (0.866667, 0.007610) (0.876190, 0.007610) (0.885714, 0.007610) (0.895238, 0.007610) (0.904762, 0.007610) (0.914286, 0.007610) (0.923810, 0.007610) (0.933333, 0.007610) (0.942857, 0.007610) (0.952381, 0.007610) (0.961905, 0.000060) (0.971429, 0.000060) (0.980952, 0.000000) (0.990476, 0.000000) (1.000000, 0.000000)
    };
     \addlegendentry{$\GreedyNaive$};
\addplot[
    color=blue,
    solid
    ]
    coordinates {
(0.000000, 967.124524) (0.009524, 902.501183) (0.019048, 810.598741) (0.028571, 810.598741) (0.038095, 642.846607) (0.047619, 642.846607) (0.057143, 617.800620) (0.066667, 617.800620) (0.076190, 543.374895) (0.085714, 543.374895) (0.095238, 479.800082) (0.104762, 467.321362) (0.114286, 429.447469) (0.123810, 429.447469) (0.133333, 288.025490) (0.142857, 106.713765) (0.152381, 101.169023) (0.161905, 91.474238) (0.171429, 75.258884) (0.180952, 48.913060) (0.190476, 41.178295) (0.200000, 24.087813) (0.209524, 24.087813) (0.219048, 3.695180) (0.228571, 3.491701) (0.238095, 2.940961) (0.247619, 1.538315) (0.257143, 1.538315) (0.266667, 0.786904) (0.276190, 0.521894) (0.285714, 0.521894) (0.295238, 0.387050) (0.304762, 0.360074) (0.314286, 0.360074) (0.323810, 0.310441) (0.333333, 0.278800) (0.342857, 0.246009) (0.352381, 0.246009) (0.361905, 0.146080) (0.371429, 0.146080) (0.380952, 0.067319) (0.390476, 0.034107) (0.400000, 0.034107) (0.409524, 0.009181) (0.419048, 0.009181) (0.428571, 0.001311) (0.438095, 0.000576) (0.447619, 0.000315) (0.457143, 0.000270) (0.466667, 0.000270) (0.476190, 0.000174) (0.485714, 0.000066) (0.495238, 0.000066) (0.504762, 0.000005) (0.514286, 0.000005) (0.523810, 0.000003) (0.533333, 0.000003) (0.542857, -0.000000) (0.552381, -0.000000) (0.561905, -0.000000) (0.571429, -0.000000) (0.580952, -0.000000) (0.590476, -0.000000) (0.600000, -0.000000) (0.609524, -0.000000) (0.619048, -0.000000) (0.628571, -0.000000) (0.638095, -0.000000) (0.647619, -0.000000) (0.657143, -0.000000) (0.666667, -0.000000) (0.676190, -0.000000) (0.685714, -0.000000) (0.695238, -0.000000) (0.704762, -0.000000) (0.714286, -0.000000) (0.723810, -0.000000) (0.733333, -0.000000) (0.742857, -0.000000) (0.752381, -0.000000) (0.761905, -0.000000) (0.771429, -0.000000) (0.780952, -0.000000) (0.790476, -0.000000) (0.800000, -0.000000) (0.809524, -0.000000) (0.819048, -0.000000) (0.828571, -0.000000) (0.838095, -0.000000) (0.847619, -0.000000) (0.857143, -0.000000) (0.866667, -0.000000) (0.876190, -0.000000) (0.885714, -0.000000) (0.895238, -0.000000) (0.904762, -0.000000) (0.914286, -0.000000) (0.923810, -0.000000) (0.933333, -0.000000) (0.942857, -0.000000) (0.952381, -0.000000) (0.961905, -0.000000) (0.971429, -0.000000) (0.980952, -0.000000) (0.990476, -0.000000) (1.000000, -0.000000)
    };
     \addlegendentry{$\Greedy\MinVar$};
\addplot[
    color=red, thick,
    dotted
    ]
    coordinates {
(0.000000, 967.124524) (0.009524, 895.801005) (0.019048, 883.322284) (0.028571, 845.448392) (0.038095, 845.448392) (0.047619, 792.777037) (0.057143, 792.777037) (0.066667, 501.619889) (0.076190, 430.812246) (0.085714, 430.812246) (0.095238, 269.760290) (0.104762, 269.760290) (0.114286, 244.714304) (0.123810, 244.714304) (0.133333, 164.743836) (0.142857, 162.124324) (0.152381, 152.429539) (0.161905, 88.854725) (0.171429, 83.739473) (0.180952, 67.524119) (0.190476, 50.433637) (0.200000, 24.087813) (0.209524, 24.087813) (0.219048, 3.695180) (0.228571, 3.491701) (0.238095, 2.940961) (0.247619, 1.538315) (0.257143, 1.300491) (0.266667, 0.786904) (0.276190, 0.521894) (0.285714, 0.506265) (0.295238, 0.477057) (0.304762, 0.457476) (0.314286, 0.457476) (0.323810, 0.357547) (0.333333, 0.357547) (0.342857, 0.222703) (0.352381, 0.145667) (0.361905, 0.145667) (0.371429, 0.096034) (0.380952, 0.065749) (0.390476, 0.034107) (0.400000, 0.034107) (0.409524, 0.008861) (0.419048, 0.008861) (0.428571, 0.001294) (0.438095, 0.000530) (0.447619, 0.000530) (0.457143, 0.000270) (0.466667, 0.000235) (0.476190, 0.000235) (0.485714, 0.000066) (0.495238, 0.000066) (0.504762, 0.000005) (0.514286, 0.000004) (0.523810, 0.000004) (0.533333, 0.000002) (0.542857, 0.000000) (0.552381, 0.000000) (0.561905, 0.000000) (0.571429, 0.000000) (0.580952, 0.000000) (0.590476, 0.000000) (0.600000, 0.000000) (0.609524, 0.000000) (0.619048, 0.000000) (0.628571, 0.000000) (0.638095, 0.000000) (0.647619, 0.000000) (0.657143, 0.000000) (0.666667, 0.000000) (0.676190, 0.000000) (0.685714, 0.000000) (0.695238, 0.000000) (0.704762, 0.000000) (0.714286, 0.000000) (0.723810, 0.000000) (0.733333, 0.000000) (0.742857, 0.000000) (0.752381, 0.000000) (0.761905, 0.000000) (0.771429, 0.000000) (0.780952, 0.000000) (0.790476, 0.000000) (0.800000, 0.000000) (0.809524, 0.000000) (0.819048, 0.000000) (0.828571, 0.000000) (0.838095, 0.000000) (0.847619, 0.000000) (0.857143, 0.000000) (0.866667, 0.000000) (0.876190, 0.000000) (0.885714, 0.000000) (0.895238, 0.000000) (0.904762, 0.000000) (0.914286, 0.000000) (0.923810, 0.000000) (0.933333, 0.000000) (0.942857, 0.000000) (0.952381, 0.000000) (0.961905, 0.000000) (0.971429, 0.000000) (0.980952, 0.000000) (0.990476, 0.000000) (1.000000, 0.000000)
    };
     \addlegendentry{$\mathsf{Best}$};
\end{axis}
\end{tikzpicture}}
  \caption{\label{fig:expr:robustness}Effectiveness of
      algorithms in reducing uncertainty in the claim robustness
      (selected datasets).}
\end{figure}

Results on robustness are similar.
  Figure~\ref{fig:expr:robustness} samples some of the results,
  specifically for \emph{CDC-firearms} and $\mathsf{UR}_x$.  Again, we
  see that \Greedy\MinVar\ and $\mathsf{Best}$ have almost the same
  performance and both outperform \GreedyNaive.  For example, for
  \emph{CDC-firearms}, using the $30\%$ of the budget, \Greedy\MinVar\
  and $\mathsf{Best}$ reduce the expected variance to almost half of \GreedyNaive.
  For $\mathsf{UR}_x$, \GreedyNaive\ performs much
  worse than \Greedy\MinVar\ and $\mathsf{Best}$.  Overall, the
  consistency of results on uniqueness and robustness is not
  surprising since our algorithms makes no assumption on the function
  used for ascertaining the claim quality (Section~\ref{sec:algos}).

\subsection{Effectiveness in Action}
\label{sec:expr:action}

\mparagraph{Ascertaining claim quality}
Earlier in this section, we have seen how our proposed algorithms
reduce the expected uncertainty in claim quality.  However, given a
specific scenario, a fact-checker using these algorithms to make data
cleaning decisions will not necessarily experience the \emph{expected}
uncertainty---instead, after the true values of cleaned objects are
revealed, a specific amount of uncertainty would remain.  To help us
evaluate the effectiveness of our algorithms in action from the
perspective of a fact-checker, we perform experiments here to simulate
a specific scenario.  We consider the same data and workload as in
Section~\ref{sec:expr:submodular}.  First, we establish the true
values for all objects, which are generated from the given value
distributions.  These values are hidden from the fact-checker and our
algorithms.  Next, as the budget varies, we let each algorithm pick
its set of objects to clean and reveal their true values, with which
we can estimate the uniqueness of the claim (in terms of the
  mean and standard deviation of the degree of duplicity).

  \begin{figure}[h]
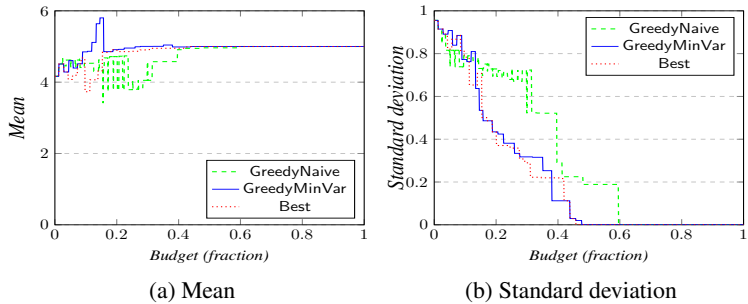

  \centering
  \subfloat[Mean]{\input{figs/expr/perc/ExUnReal}}
  \subfloat[Standard deviation]{\input{figs/expr/perc/VarUReal}}
  \caption{\label{fig:Var}Mean and standard deviation of
      estimates of claim uniqueness as functions of budget.
      \emph{CDC-causes}.}
\end{figure}

  Figure~\ref{fig:Var} plots the mean and standard deviation
  (respectively) of the estimates resulted from each algorithm's
  decision as functions of budget. The dataset here is
  \emph{CDC-causes}, and the claim has the same form as in
  Figure~\ref{fig:expr:un:real}b.  For this specific scenario, the
  true degree of duplicity for the claim happens to be $5$.  From
  Figure~\ref{fig:Var}, we see that $\mathsf{Best}$ and
  \Greedy\MinVar\ generate better estimates faster than \GreedyNaive.
  For example, at $20\%$ of the total cost, \GreedyNaive's estimated
  mean is $3.7$, with a standard deviation of $0.7$, which is still
  difficult for the fact-checker to gauge true uniqueness.  In
  comparison, \Greedy\MinVar\ and $\mathsf{Best}$ finds the mean to be
  $4.9$ (which is closer to the true value) with a lower standard
  deviation of $0.4$.

  \begin{figure}[h]
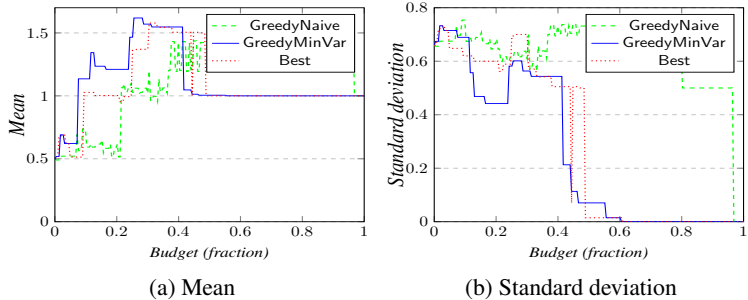

  \centering
  \subfloat[Mean]{\input{figs/expr/perc/uni-unic-r100-EM}}
  \subfloat[Standard deviation]{\input{figs/expr/perc/uni-unic-r100-Var}}
  \caption{\label{fig:Var:synthetic}Mean and standard deviation of claim uniqueness
    as functions of budget for a specific scenario with value
    distributions from $\mathsf{UR}_x$ and $\Gamma=100$.}
\end{figure}

We run the same experiment with synthetic datasets.
Figures~\ref{fig:Var:synthetic}a
and~\ref{fig:Var:synthetic}b plot the mean and standard deviation (respectively) of
claim uniqueness resulted from each algorithm's decision as functions
of budget. The value distributions here are from $\mathsf{UR}_x$ and
the parameter $\Gamma$ in the query function is set to $100$; other
distributions and $\Gamma$ values give similar results and are hence
omitted.  For this specific scenario, the true uniqueness of the claim
happens to be $1$.  From Figure~\ref{fig:Var}, we see that
$\mathsf{Best}$ and \Greedy\MinVar\ perform again much better than
\GreedyNaive.  For example, at $20\%$ of the total cost, \GreedyNaive\
finds the mean to be $0.5$ with a standard deviation of $0.63$; it
is still difficult to the fact-checker to gauge true uniqueness.  In
comparison, \Greedy\MinVar\ finds the expected uniqueness to be $1.2$
(which is close to the true uniqueness) with a lower variance of
$0.44$.


  Generally speaking, combining the results here and
  those in Section~\ref{sec:expr:submodular}, we observe that
  \Greedy\MinVar\ in expectation requires cleaning less data than
  \GreedyNaive\ for fact-checkers to assess claim qualities.

\mparagraph{Finding counters}
Similarly, we simulate scenarios to evaluate how our algorithms can
help find counterarguments. We describe the results for one
  scenario on \emph{CDC-firearms} and $\mathsf{UR}_x$ (with $\Gamma=100$) datasets.
  To establish the (hidden) true
  values as well as the current (noisy) values, we randomly sample
  from the value distribution of each object.

  For the \emph{CDC-firearms} we want to check the
  claim that ``in the past four year, we had only 310000 injuries by
  firearms, lowest in recent history.''  If we assume the current
  noisy values to be correct, there would be no counterexample in the
  database, i.e., there is no other period with fewer injuries.
  However, if we clean all data to reveal the true values, there is a
  counterargument for the period 2002--2006.  A fact-checker must
  clean some tuples to counter the original claim.  We observe that
  \Greedy\MaxPr\ uses only $7\%$ of the budget to find the
  couterargument with high probability (more than $98\%$), while
  \GreedyNaive\ uses $74\%$ of the budget to achieve the same.

  Similarly for $\mathsf{UR}_x$ there
is no counter to the original claim assuming current object values;
however, with true object values, there indeed exists a strong counter
to the original claim.  In other words, a fact-checker must clean some
values in order to find a counter.
After cleaning only $3$ values and using $8\%$ of the total cost,
\Greedy\MaxPr\ is able to find a counter to the original claim.  In
comparison, \GreedyNaive\ is much less effective: it finds a counter
only after cleaning $15$ values and using $21\%$ of the total cost.


\subsection{Efficiency}
\label{sec:expr:efficiency}

Having seen the effectiveness of \Greedy\MinVar\ in earlier
experiments, we now evaluate its efficiency.  We note that
$\mathsf{Best}$ is generally slower than \Greedy\MinVar\ (often by
factors of more than $5$ in our experiments) and does not seem to
deliver better solutions in practice.  On the other hand,
\GreedyNaive\ is much faster than \Greedy\MinVar\ because of naive
benefit estimation, but it is not nearly as effective as shown
earlier.

First, we consider the same scenarios as in
Figure~\ref{fig:expr:uniqueness:univ-unic}, but scale up each synthetic dataset
to contain $10{,}000$ uncertain values.  We also proportionally
increase the number of perturbations considered to $2{,}500$ such that
together they cover all values.  We report the results for
$\mathsf{UR}_x$; other datasets are similar.
Figure~\ref{fig:expr:time:univ-unic}a shows the running time of
\Greedy\MinVar\ as we give it increasing budgets to work with.  We see
that running time increases roughly linearly with budget.  Even with a
budget that allows $30\%$ of all data to be cleaned, \Greedy\MinVar\
completes under $2$ minutes.

Next, to study the effect of dataset size on running time, we consider
progressively bigger datasets, from $50{,}000$ to $1{,}000{,}000$
uncertain values, whose distributions are still generated using
$\mathsf{UR}_x$.  Again, we scale up the number of perturbations
considered accordingly.  We fix the budget at $5000$ to allow about
$1{,}000$ values to be cleaned.  Figure~\ref{fig:expr:time:univ-unic}b
shows how \Greedy\MinVar's running time (in $\log_{10}$ scale)
increases with data size.  We observe that each time that the data
size increases by a factor of $10$, the running time to clean about
$1{,}000$ tuples is $18$--$19$ times larger.  Even with a large
dataset containing $100{,}000$ uncertain values, it takes less than
$12$ minutes for \Greedy\MinVar\ to suggest cleaning $1{,}000$ values,
which translates to about $0.725$ seconds per recommendation.

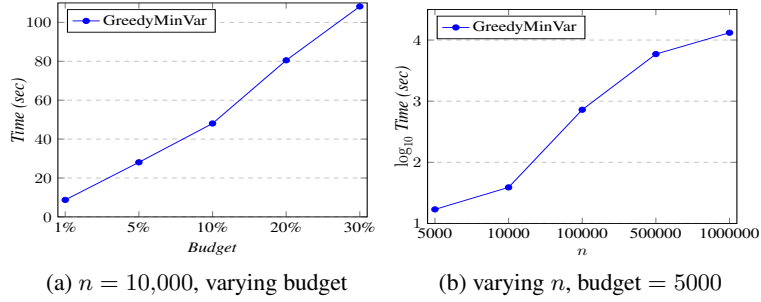
\begin{figure}[t]
\centering
  \subfloat[$n = 10{,}000$, varying budget]{\begin{tikzpicture}[yscale=0.5, xscale=0.6]
\begin{axis}[
    y label style={at={(axis description cs:0.1,.5)},anchor=south},
    xlabel=\emph{Budget},
    ylabel=\emph{{\large Time (sec)}},
    xmin=0.9, xmax=5.1,
    ymin=0, ymax=110,
    xtick={1, 2, 3, 4, 5},
    xticklabels={1\%, 5\%, 10\%, 20\%, 30\%},
    legend pos=north west,
    ymajorgrids=true,
    grid style=dashed,
]

\addplot[
    color=blue,
    mark=*,
    ]
    coordinates {
    (1, 8.737) (2, 28.057) (3, 48.02) (4, 80.498) (5, 108.186)
    };
     \addlegendentry{GreedyMinVar};
\end{axis}
\end{tikzpicture} }
  \subfloat[varying $n$, budget $=5000$]{\begin{tikzpicture}[yscale=0.5, xscale=0.6]
\begin{axis}[
    y label style={at={(axis description cs:0.15,.5)},anchor=south},
    xlabel=\emph{$n$},
    ylabel=\emph{{\large $\log_{10}$ Time (sec)}},
    xmin=0.9, xmax=5.1,
    ymin=1, ymax=4.5,
    xtick={1, 2, 3, 4, 5},
    xticklabels={5000, 10000, 100000, 500000, 1000000},
    legend pos=north west,
    ymajorgrids=true,
    grid style=dashed,
]

\addplot[
    color=blue,
    mark=*,
    ]
    coordinates {
     (1, 1.23) (2, 1.59) (3, 2.86) (4, 3.77) (5, 4.12)
    };
     \addlegendentry{GreedyMinVar};
\end{axis}
\end{tikzpicture} }
  \caption{\label{fig:expr:time:univ-unic}Running time of
    \Greedy\MinVar\ when reducing uncertainty in claim uniqueness.
    $\mathsf{UR}_x$; claim with $\Gamma=100$.}
\end{figure}

\subsection{Handling Dependency}
\label{sec:expr:dependency}

We note that our theoretical guarantees require the
  independence assumption among the object values.  Nevertheless, our
  algorithms can still be applied to situations where the independence
  assumption does not hold.  To see how our algorithms perform
  practically in such situations, we design additional experiments by
  modifying the \emph{CDC-firearms} dataset.  Recall that for each
  object $X_i$, CDC reports its standard deviation $\sigma_i$.
  Although errors across $X_i$'s are actually independent because of
  CDC's data sampling procedure, we artificially introduce dependency
  as follows.  We create a covariance matrix where the covariance
  between two objects $X_i, X_j$ (where $i<j$ refer to the years) is
  given by $\gamma^{j-i}\sigma_i\sigma_j$, where the parameter
  $\gamma \in [0,1]$ controls the degree of the dependency (the closer
  $\gamma$ is to $1$, the more dependent $X_i$'s become).  The
  exponent $j-i$ in this covariance model captures the intuition that
  the farther apart the two years are, the smaller their dependency
  is.  The claim in question is the same as in
  Section~\ref{sec:expr:modular}.

Since we do not have an efficient algorithm capable of
  handling dependencies with good theoretical guarantees, for this
  experiment, where the dataset is thankfully small, we resort to a
  brute-force algorithm $\mathsf{OPT}$ and use it as a yardstick for
  comparison.  $\mathsf{OPT}$ has full knowledge of data dependency
  (i.e., the covariance matrix), exhaustively considers all possible
  subsets of values to clean, and returns the best subset satisfying
  the cost constraint.  $\mathsf{Optimum}$, \Greedy\MinVar,
  \GreedyNaive, and $\GreedyNaive\mathsf{CostBlind}$ are not made
  aware of any dependency at all.  In addition, we implement a variant
  of \Greedy\MinVar\ called $\Greedy\mathsf{Dep}$, which is given the
  dependency knowledge and uses it for estimating cleaning benefits.

\begin{figure}[h]
\centering
  \subfloat[$\gamma=0.7$, varying budget]{\input{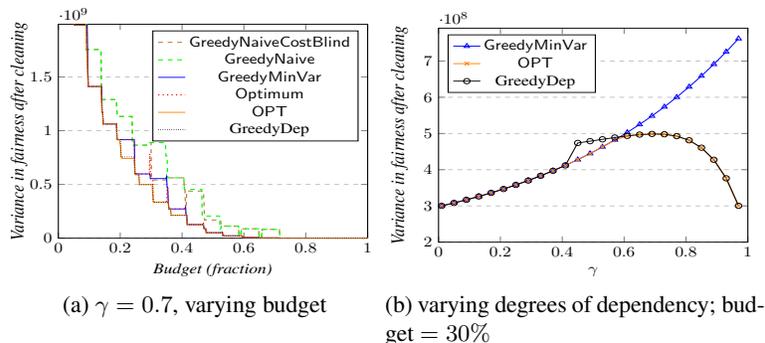}}
  \subfloat[varying degrees of dependency; budget $=30\%$]{\begin{tikzpicture}[yscale=0.5, xscale=0.6]
\begin{axis}[
    y label style={at={(axis description cs:0.1,.5)},anchor=south},
    xlabel=\emph{$\gamma$},
    ylabel=\emph{\large{Variance in fairness after cleaning}},
    xmin=0.0, xmax=1.0,
    ymin=200000000, ymax=790814720,
    legend pos=north west,
    ymajorgrids=true,
    grid style=dashed,
]
\addplot[
    color=blue,
    mark=triangle
    ]
    coordinates {
(0.010000, 300141461.228162) (0.050000, 308226508.490418) (0.090000, 316878695.630638) (0.130000, 326138340.121053) (0.170000, 336048099.849421) (0.210000, 346653090.165786) (0.250000, 358001004.581019) (0.290000, 370142237.180612) (0.330000, 383130004.847049) (0.370000, 397020467.414495) (0.410000, 411872843.909501) (0.450000, 427749523.060102) (0.490000, 444716166.281646) (0.530000, 462841801.369611) (0.570000, 482198905.145778) (0.610000, 502863473.312662) (0.650000, 524915075.769859) (0.690000, 548436895.632799) (0.730000, 573515750.166660) (0.770000, 600242091.803339) (0.810000, 628709987.344349) (0.850000, 659017073.364335) (0.890000, 691264485.715104) (0.930000, 725556760.885277) (0.970000, 762001706.792010)
        };
     \addlegendentry{\Greedy\MinVar};

\addplot[
    color=orange,
    mark=x
    ]
    coordinates {
(0.010000, 300141461.228162) (0.050000, 308226508.490418) (0.090000, 316878695.630638) (0.130000, 326138340.121053) (0.170000, 336048099.849421) (0.210000, 346653090.165786) (0.250000, 358001004.581019) (0.290000, 370142237.180612) (0.330000, 383130004.847049) (0.370000, 397020467.414495) (0.410000, 411872843.909501) (0.450000, 427749523.060102) (0.490000, 444716166.281646) (0.530000, 462841801.369611) (0.570000, 482198905.145778) (0.610000, 493355848.565629) (0.650000, 496902577.772689) (0.690000, 498742590.442833) (0.730000, 497822500.722861) (0.770000, 492668106.223281) (0.810000, 481241904.968402) (0.850000, 460757334.565668) (0.890000, 427438070.607306) (0.930000, 376208042.946650) (0.970000, 300294642.172785)
    };
     \addlegendentry{$\mathsf{OPT}$};

\addplot[
    color=black,
    mark=o
    ]
    coordinates {
(0.010000, 300141461.228162) (0.050000, 308226508.490418) (0.090000, 316878695.630638) (0.130000, 326138340.121053) (0.170000, 336048099.849421) (0.210000, 346653090.165786) (0.250000, 358001004.581019) (0.290000, 370142237.180612) (0.330000, 383130004.847049) (0.370000, 397020467.414495) (0.410000, 411872843.909501) (0.450000, 473834635.567633) (0.490000, 478800098.887869) (0.530000, 483877171.843935) (0.570000, 488843599.984854) (0.610000, 493355848.565629) (0.650000, 496902577.772689) (0.690000, 498742590.442833) (0.730000, 497822500.722861) (0.770000, 492668106.223281) (0.810000, 481241904.968402) (0.850000, 460757334.565668) (0.890000, 427438070.607306) (0.930000, 376208042.946650) (0.970000, 300294642.172785)
    };
    \addlegendentry{GreedyDep};
\end{axis}
\end{tikzpicture} }
  \caption{\label{fig:expr:NonIndF}Effectiveness of
      algorithms in reducing uncertainty in claim fairness.
      \emph{CDC-firearms} with dependencies injected.}
\end{figure}

Figure~\ref{fig:expr:NonIndF}a compares how effective these
  algorithms are in reducing uncertainty in claim fairness, with
  parameter $\gamma$ controlling the degree of dependency set to
  $0.7$.  We see that $\mathsf{Optimum}$ and \Greedy\MinVar\ always
  perform better than the simpler \GreedyNaive\ and
  $\GreedyNaive\mathsf{CostBlind}$.  Furthermore, in many cases
  $\mathsf{Optimum}$ and \Greedy\MinVar\ have the same efficacy as
  $\mathsf{OPT}$.  For example, using only $18\%$ budget,
  $\mathsf{Optimum}$ and \Greedy\MinVar\ have already reduced the
  uncertainty to less than half of the initial uncertainty, matching
  the performance of $\mathsf{OPT}$.  Of course, as expected, having
  the knowledge of data dependency helps---there are cases where
  $\mathsf{Optimum}$ or \Greedy\MinVar\ still fail to match the
  performance of $\mathsf{OPT}$.  Interestingly, once given the
  knowledge of data dependency, $\Greedy\mathsf{Dep}$, despite still
  being greedy, almost always matches the performance of
  $\mathsf{OPT}$ in this case.  Overall, we observe that knowing the
  dependencies is beneficial, and the same greedy strategy is capable
  of reaping much of this benefit.  Even if dependencies exist but are
  not known, \Greedy\MinVar\ and $\mathsf{Optimum}$ are still viable
  practical solutions for moderate degrees of dependency.

Figure~\ref{fig:expr:NonIndF}b compares the effectiveness of
  algorithms when we vary the degree of data dependency, while fixing
  budget at $30\%$.  If the dependency is weak enough, namely when
  $\gamma \leq 0.6$, then \Greedy\MinVar, even
  though it is not aware of any dependency, performs optimally. As
  dependency grows stronger, however,\\
  \Greedy\MinVar\ starts to fall behind $\mathsf{OPT}$.  Again,
  interestingly, $\Greedy\mathsf{Dep}$ almost always matches the
  performance of $\mathsf{OPT}$, except for a small range of
  ``middle'' $\gamma$ values, where intuitively the problem is the
  hardest (in contrast, having either independent values or highly
  correlated values makes it easier to resolve uncertainty).  Overall,
  we conclude that even without any knowledge of dependency,
  \Greedy\MinVar\ (and $\mathsf{Optimum}$) are viable as long as the
  degree of dependency is not too high.  However, strong data
  dependencies would require an algorithm aware of such dependencies,
  but the greedy strategy is still effective.

\subsection{Competing Objectives}
\label{sec:expr:diff-objectives}

Theorem~\ref{theorem:minvar-maxpr-equivalence} shows how the
objectives of ascertaining claim fairness and increasing the chance of
finding counters can be aligned if errors in data are normal and
centered at $0$.  Here, we design experiments to show how the two
objectives lead to very different outcomes when this assumption does
not hold.

We return to the adoption scenario of Section~\ref{sec:expr:modular},
but simplify the claim to be about the sum over a $4$-year window and
consider perturbations with non-overlapping windows (same as in
Section~\ref{sec:expr:submodular}).  Recall that
Theorem~\ref{theorem:minvar-maxpr-equivalence} applies if all
distributions are normal and centered around the current values. Here,
we instead reassign the current values to random draws from these
distributions, so they can deviate from the mean of the respective
distributions.

\begin{figure}[h]
\centering
  \subfloat[reducing uncertainty][reducing uncertainty\newline(objective of \MinVar)]{\input{figs/expr/perc/diff-obj-var}}
  \subfloat[improving chance of countering][improving prob.\ of countering\newline(objective of \MaxPr)]{\input{figs/expr/perc/diff-obj-pr}}
  \caption{\label{fig:expr:diff-obj}How $\mathsf{Optimum}$ (for
    \MinVar) and \Greedy\MaxPr\ (for \MaxPr) achieve different
    objectives.}
\end{figure}

For ascertaining fairness, we use $\mathsf{Optimum}$ as described in
Section~\ref{sec:expr:modular}, which always finds the optimum
solution.  For maximizing the chance of finding counters, we use
\Greedy\MaxPr. We run these two algorithms, but we also
  measure how they perform with respect to the objective that they are
  \emph{not} optimizing for. Figure~\ref{fig:expr:diff-obj}a
compares how the two algorithms achieve the objective of ascertaining
claim fairness, while Figure~\ref{fig:expr:diff-obj}b compares how
they achieve the objective of maximizing the chance of finding
counters.  The vertical axes show the respective objective function
values, which are computed given the choices made by the
algorithms.\footnote{\small Note that the dataset's current values do
  not affect uncertainty in claim fairness, but do affect the chance
  of finding counters significantly.  Hence, for
  Figure~\ref{fig:expr:diff-obj}b, we repeat each experiment $100$
  times with different random draws of the current values, and report
  the average.}

From Figure~\ref{fig:expr:diff-obj}, as expected, each algorithm does
well in terms of its intended objective.  What is more revealing is
how poorly they do with regard to the other objective.  While
$\mathsf{Optimum}$ quite effectively reduces the uncertainty in claim
fairness (Figure~\ref{fig:expr:diff-obj}a), it does not offer a good
chance of finding counters even when given generous budgets
(Figure~\ref{fig:expr:diff-obj}b).  On the other hand, \Greedy\MaxPr\
quickly increases the chance of finding counters
(Figure~\ref{fig:expr:diff-obj}b), but it is much less helpful in
ascertaining claim fairness (Figure~\ref{fig:expr:diff-obj}a).  In
fact, when given a budget above $48\%$ of the total cost of cleaning
all data, \Greedy\MaxPr\ simply refuses to clean any more values
because doing so would actually decrease the chance of finding a
counter (which explains why its achieved objective function values
stay flat beyond this point).  This questionable behavior illustrates
the danger of a utilitarian approach of cleaning data for
fact-checking that just seeks to maximize the chance of countering a
claim.

\section{Related Work}
\label{sec:related}

The fact-checking aspect of this paper builds on the
framework in~\cite{wu2014toward, tods17-WuEtAl-comp_fact_check}.
However, that framework assumed an accurate database and did not
consider data cleaning.

There is a rich body of literature on data cleaning;
see~\cite{DBLP:journals/ftdb/IlyasC15,
  DBLP:series/synthesis/2013Ganti} for surveys.  A number of papers
have addressed the specific problem of cleaning data under a budget
constraint, with the goal of improving query quality.  Cheng et
al.~\cite{cheng2008cleaning} proposed a metric for query result
quality called \emph{PWS-quality}, together with efficient algorithms
for handling range and max queries.  Mo et al.~\cite{mo2013cleaning}
further tackled top-$k$ queries.  The PWS-quality is based on entropy,
and has nice properties for filter and ranking queries that, together
with an independence assumption, result in modular optimization
objectives.  In contrast, our measure of uncertainty is based on
expected variance; it is more suitable than entropy for numeric
results, which arise naturally in the application of fact-checking.
Our query functions are generally far more complex, which lead to
non-modular, more difficult optimization problems.  We also consider
the alternative objective of maximizing surprise.

There are other models of data cleaning with different goals.  For
example, \emph{ActiveClean}~\cite{krishnan2016activeclean} proposed
interactive data cleaning for statistical modeling.  Given a budget,
its goal was to choose a sample of data to clean while preserving
provable convergence properties of stochastic gradient descent.
\emph{SampleClean}~\cite{wang2014sample} used sampling to clean data
in order to improve the quality of aggregate results by minimizing the
impact of dirty data. It focuses more on improving estimation than on
picking which value to clean.
In \cite{dolatshah2018cleaning} the authors propose \emph{TARS}, which is a label cleaning advisor
that provides valuable information when a model is trained or tested using noisy labels.
They describe a model evaluation and a cleaning strategy: TARS decides which labels in
the training set should be cleaned to improve the performance of the model.
They also use the notion of \emph{expected model improvement} to define their cleaning strategy which is related to
our expected variance.
While these approaches are also
stochastic, their goals and technical challenges are quite different
from ours.
In addition, HoloClean~\cite{rekatsinas2017holoclean} focuses on automatic repairing the whole dataset, given both
constraints and known statistical properties of the input data.  It
first uses a process to detect noisy values, and then use a factor
graph specified using inference rules to model the joint distribution
of values; clean values are treated as labeled examples to learn the
model parameters, and then noisy values are assigned their maximum
a-posteriori estimates.  HoloClean is not directly applicable to our
setting because our goal is not automatic repair of the whole dataset,
but instead, selective cleaning of particular values to help
fact-check a given claim.  Also, relying on automatically derived
point estimates for fact-checking will likely be unconvincing in
practice, given that claims are often controversial.
Generally, we cannot directly compare with most of the
  data cleaning techniques that focus on automatic repairs given
  constraints~\cite{DBLP:journals/ftdb/IlyasC15,
    DBLP:series/synthesis/2013Ganti} since  we focus on the orthogonal
  problem of selecting which data items to clean;
  how to clean the particular data items is a separate
  concern that we do not address in this paper.

Kanagal et al.~\cite{DBLP:conf/sigmod/KanagalLD11} studied how to
quantify the \emph{influence} of input tuples on the results of
queries over a probabilistic database.  Our work is related in the
general sense that we seek to quantify the benefit of cleaning a value
given an objective function defined using some query of
interest.
However, their work is quite different from ours in both the
uncertainty model and functions
they support (which are less complex).

There is also a large body of literature on approximation algorithms
for stochastic data~\cite{li2015approximation, dean2004approximating,
  steinberg1979preference, morton1998stochastic, ilhan2011technical}.
Particularly related is the line of work on sensing---how to place
sensors or probe data in order to maximize utility under a budget
constraint.  Krause et al.~\cite{krause2008near} considered the
problem of placing sensors in order to maximize the \emph{mutual
  information}.  Their objective function is submodular, so a greedy
algorithm achieves a $(1-1/e)$-approximation.  Our technical
challenges are different because our objective function is not
entropy-based, and the direction of optimization is also different
(minimization vs.\ maximization), which turns out to be crucial (more
on this point below).

In~\cite{krause2008robust}, Krause et al.~studied the submodular
observation selection problem.  The goal is to select a subset of
locations $T \subseteq \Objs$ to observe in order to minimize the
average \emph{predictive variance}. Specifically, they aim to find a
subset $T$ with $|T|\leq k$, to minimize
$V(T)=\frac{1}{n}\sum_i\int\Prob{\*X_T}\Expected{X_i-\Expected{X_i
    \mid\*X_T})^2\mid\*X_T} d\*X_T$.  This notion of predictive
variance is similar to our expected variance.  However, the problems
are quite different.  They found that the function $V(T)$ is
supermodular, or equivalently, the \emph{variance reduction} function
$V(\emptyset)-V(T)$ is submodular.  However, we found that our
$\EVar{T}$ is submodular (under the conditions in Lemma
\ref{lemma:submodular}), or equivalently, the variance reduction
function $\EVar{\emptyset}-\EVar{T}$ is supermodular, which is the
exact opposite of the property they derived and requires a different
approach.  There is no inconsistency, however: their problem
corresponds to the case where (in our terminology) the query function
is linear but errors may be correlated; while our result is for the
case where the errors are mutually independent but the query function
can be arbitrarily complex.  The only case where the two problems
become the same is when the query function is linear and errors are
mutually independent, which means the objective function is modular
and both problems reduce to the simpler knapsack problem.



\section{Conclusion and Future Work}
\label{sec:conclude}

In this paper, we have considered how to help fact-checkers combat the
issues of data quality and data fishing, by combining data cleaning
and perturbation analysis, and by solving the optimization problem of
choosing a subset of data to clean under a budget constraint, with the
goal of either minimizing uncertainty in claim quality or maximizing
the chance of find counters.  We have demonstrated through experiments
that our proposed algorithms are effective and efficient in practice.  We
also have shown when the two optimization goals align and when they do
not.  In sum, our results provide practical tools and guidelines that
help fact-checkers clean data effectively while avoiding the potential
bias introduced by their eagerness to counter claims.

There are several interesting directions for future work.  First,
better algorithms may be needed to solve the optimization problem for
arbitrary query functions in the presence of correlated errors.
Second, instead of making all choices upfront, an algorithm can adapt
its data cleaning actions to the outcome of its earlier actions, which
is particularly useful to \MaxPr.  Finally, it will be useful to study
settings where cleaning an individual value only reduces the
uncertainty thereof, but does not completely eliminate it.


\bibliographystyle{abbrv}
\bibliography{counterClaims}
\newpage
\section{Appendix}

\subsection{Missing proofs for modular objectives}
\label{appndx:modular}

\begin{proof}{ of Lemma~\ref{lemma:linear-modular}}
  We first show the first part of the lemma.
  Note that
  \begin{align*}
    \Var{f(\*X) \mid \*X_T = \*v}
    & = \Var{b + \sum_{\obj_i \in T} a_i v_i + %
      \sum_{\obj_i \in \Objs \setminus T}  a_i X_i} \\%
    &  = \Var{\sum_{\obj_i \in \Objs \setminus T} a_i X_i}\\ %
    &  = \sum_{\obj_i \in \Objs \setminus T}%
      \,\sum_{\obj_j \in \Objs \setminus T}%
      \Cov{a_iX_i, a_jX_j}.
  \end{align*}
  Since components of $\*X$ are pairwise uncorrelated,
  $$\Cov{a_iX_i, a_jX_j}=0$$ for all $i \neq j$.
  Therefore, the above sum simplifies to
  $$\sum_{\obj_i \in \Objs \setminus T} \Var{a_iX_i} = \sum_{\obj_i \in
    \Objs \setminus T} a_i^2 \Var{X_i}.$$
  Hence, minimizing\\$\sum_{\*v \in \*V_T}
  \Prob{\*X_T = \*v} \cdot \Var{f(\*X) \mid \*X_T = \*v}$
  is the same as minimizing\\
  $\sum_{\*v \in \*V_T} \left(%
      \Prob{\*X_T = \*v} \cdot%
      \sum_{\obj_i \in \Objs \setminus T}  a_i^2 \Var{X_i}%
      \right)
      = \left(\sum_{\*v \in \*V_T} \Prob{\*X_T = \*v}\right) \cdot%
      \left(\sum_{\obj_i \in \Objs \setminus T} a_i^2 \Var{X_i}\right)
      = \sum_{\obj_i \in \Objs \setminus T} a_i^2 \Var{X_i}$.
Equivalently we can write the problem as minimizing
$\sum_{\obj_i \in T'} a_i^2 \Var{X_i}$ such that $\sum_{\obj_i\in T'}c_i\geq \sum_{\obj_i\in \Objs}c_i-C$.
  Therefore, the objective is
  additive, with $w_i = a_i^2 \Var{X_i}$.

In order to prove the second part of the lemma,
  note that if $\*X_{\Objs \setminus T} = \*u_{\Objs \setminus T}$,
  then
  \begin{align*}
    f(\*X) - f(\*u) &= %
    (b + \*a_T\*X_T +%
    \*a_{\Objs \setminus T}\*u_{\Objs \setminus T}) -%
(b + \*a_T\*u_T +%
    \*a_{\Objs \setminus T}\*u_{\Objs \setminus T})\\ %
    &= \*a_T ( \*X_T - \*u_T ).
  \end{align*}
  Since $X_i \sim N(u_i, \sigma_i^2)$ and all $X_i$'s are independent,
  $f(\*X) - f(\*u) \sim N(0, \sum_{\obj_i \in T} a_i^2 \sigma_i^2)$.
  Therefore,
  \begin{align*}
    \Prob{f(\*X) < f(\*u) - \tau \mid \*X_{\Objs \setminus T} = \*u_{\Objs \setminus T}} %
   &= \Prob{N(0, \sum_{\obj_i \in T} a_i^2 \sigma_i^2) < -\tau}\\ %
   & = \Prob{N(0, 1) < %
    {-\tau \over \sqrt{\sum_{\obj_i \in T} a_i^2 \sigma_i^2}}}.
  \end{align*}
  Maximizing the above is equivalent to maximizing
  $\sum_{\obj_i \in T} a_i^2 \sigma_i^2$, so the objective is additive
  with $w_i = a_i^2 \sigma_i^2$.
\end{proof}

\begin{proof}{ of Lemma~\ref{corollary:modular-maxpr}}
From Lemma~\ref{lemma:linear-modular} we can formulate the \MaxPr\ problem as follows.
$$\max_{T\subseteq \Objs, \sum_{\obj_i\in T} c_i\leq C}\Prob{\sum_{\obj_i\in T}a_iX_i< -\tau},$$
where $a_i$ are real numbers and $\tau$ is a positive number.
Following the analysis of \cite{morton1998stochastic} we can write our objective as
\begin{align*}
   \Prob{\sum_{\obj_i\in T}a_iX_i< -\tau}
  &= \Prob{\sum_{\obj_i\in T} a_i\cdot \Expec{X_i}+ \mathcal{N}(0,1)\sqrt{\sum_{\obj_i\in T} a_i^2\cdot \Var{X_i}}<-\tau}\\
						&= \Prob{\mathcal{N}(0,1)\sqrt{\sum_{\obj_i\in T} a_i^2\cdot \Var{X_i}}< -\tau}\\
						&= \Prob{\mathcal{N}(0,1)< \frac{-\tau}{\sqrt{\sum_{\obj_i\in T} a_i^2\cdot \Var{X_i}}}}
\end{align*}
The maximum probability above is taken when the quantity\\$\frac{-\tau}{\sqrt{\sum_{\obj_i\in T} a_i^2\cdot \Var{X_i}}}<0$
is maximized. So the optimum subset to clean is the same with the optimum subset of the following problem:
\[
\bar{P}: \max_{T\subseteq \Objs, \sum_{\obj_i\in T}c_i\leq C}\sum_{\obj_i\in T}a_i^2\cdot \Var{X_i}
\]
Problem $\bar{P}$ is an instance of the Knapsack problem, so using
the exact dynamic pseudo-polynomial algorithm we can get an exact solution.
Intuitively, it says that if all $X_i$'s have mean equals to $0$,
then we should choose to clean objects with high variance and small cost in order to
maximize the probability that $\sum_{\obj_i\in T}a_iX_i< -\tau$.
The first part of Lemma~\ref{corollary:modular-maxpr} follows.

For the second part of the Lemma, in order to get an approximation algorithm,
we use an FPTAS for $\bar{P}$ (because $\bar{P}$ is an instance of the Knapsack problem).
Let $P^*$ be the value of the optimum solution for $\bar{P}$ and $A$ the solution of the FPTAS.
We have that $A\geq (1-\epsilon)P^*$.
By doing simple calculations we have $\frac{-\tau}{\sqrt{A}}\geq \frac{1}{\sqrt{1-\epsilon}}\frac{-\tau}{\sqrt{P^*}}$.
Let $P^*_1 = \frac{-\tau}{\sqrt{P^*}}$ and $A_1=\frac{-\tau}{\sqrt{A}}$. We assume that $\epsilon= 3/4$ and we have
\[
 A_1\geq 2\cdot P^*_1.
\]
The maximum probability is computed by $\Prob{\mathcal{N}(0,1)< P^*_1}$.
In our case we have,
\[
 \Prob{\mathcal{N}(0,1)< A_1}\geq \Prob{\mathcal{N}(0,1)< 2\cdot P^*_1}
\]
The overall approximation error would be the value of the minimum possible ratio
\[
\frac{\Prob{\mathcal{N}(0,1)< 2\cdot P^*_1}}{\Prob{\mathcal{N}(0,1)< P^*_1}}=
\frac{\int_{2\cdot P^*_1}^{-\infty}e^{-\frac{1}{2}t} \diff t}{\int_{P^*_1}^{-\infty}e^{-\frac{1}{2}t} \diff t}
\]
Unfortunately, the function $\frac{\int_{2\cdot x}^{-\infty}e^{-\frac{1}{2}t} \diff t}{\int_{x}^{-\infty}e^{-\frac{1}{2}t} \diff t}$
is decreasing as $x<0$ and
$\lim_{x\rightarrow -\infty}\frac{\int_{2x}^{-\infty}e^{-\frac{1}{2}t} \diff t}{\int_{x}^{-\infty}e^{-\frac{1}{2}t} \diff t}=0$,
so the approximation ratio can be arbitrarily small, for very small values of $P^*_1$.
Notice that if $P^*_1$ is small, then $\Prob{\mathcal{N}(0,1)< P^*_1}$
is very small. For example, even if $P^*_1=-4$, then $\Prob{\mathcal{N}(0,1)< P^*_1}=0.00003$.
In these cases it is usual to consider that there is a lower bound on the minimum possible probability.
In practice if the maximum probability is very small then we can consider it zero.
In our case we assume that, if $\Prob{\mathcal{N}(0,1)< P^*_1}<0.05$
then we can safely consider that the probability is equal to $0$. If we set,
$$\Prob{\mathcal{N}(0,1)< P^*_1} = \int_{P^*_1}^{-\infty}e^{-\frac{1}{2}t} \diff t= 0.05$$ then
we find $P^*_1\approx -1.64$, and so we assume that $P^*_1\geq -1.64$ in order not to have a $0$
probability. The worst approximation ratio is:
\[
 \frac{\int_{2\cdot (-1.64)}^{-\infty}e^{-\frac{1}{2}t} \diff t}{\int_{-1.64}^{-\infty}e^{-\frac{1}{2}t} \diff t} \geq 1/100=O(1).
\]
The second part of Lemma~\ref{corollary:modular-maxpr} follows.
\end{proof}




\subsection{Missing proofs for general query functions}
\label{appndx:proofsSub}

\begin{lemma}\label{lemma:power-mean}
  Suppose $\sum_{i=1}^n w_i = 1$, and $0 \le w_i \le 1$ for
  all $1 \le i \le n$.  Then
  \begin{align*}
    \left(\sum_{i=1}^n w_ix_i^2\right) - %
    \left(\sum_{i=1}^n w_ix_i\right)^2%
    = \sum_{1 \le i < j \le n} w_iw_j(x_i-x_j)^2%
    \ge 0.
  \end{align*}
\end{lemma}
\begin{proof}{}
  Note that the first summation is the square of the generalized
  weighted power mean of exponent $2$, while the second summation is
  the square of the generalized weighted power mean of exponent $1$.
  \begin{align*}
    \left(\sum_{i=1}^nw_ix_i^2\right) - %
    \left(\sum_{i=1}^n w_ix_i\right)^2%
   &= \left(\sum_{i=1}^n w_ix_i^2\right) - %
      \left(\sum_{i=1}^n w_i^2x_i^2\right) - %
      2\sum_{1 \le i < j \le n} w_ix_iw_jx_j\\
    & = \left(\sum_{i=1}^n w_ix_i^2(1-w_i)\right) - %
      2\sum_{1 \le i < j \le n} w_ix_iw_jx_j\\
    & = \left(\sum_{i=1}^n w_ix_i^2 %
      \Big(\sum_{j \neq i, 1 \le j \le n} w_j\Big)\right) - %
      2\sum_{1 \le i < j \le n} w_ix_iw_jx_j\\
    & = \left(\sum_{i \neq j, 1 \le i,j \le n} w_iw_jx_i^2\right) - %
      2\sum_{1 \le i < j \le n} w_ix_iw_jx_j\\
    & = \left(\sum_{1 \le i < j \le n} w_iw_j(x_i^2+x_j^2)\right) - %
      2\sum_{1 \le i < j \le n} w_ix_iw_jx_j\\
    & = \sum_{1 \le i < j \le n} w_iw_j(x_i^2+x_j^2-2x_ix_j)\\
    & = \sum_{1 \le i < j \le n} w_iw_j(x_i-x_j)^2%
      \ge 0.
  \end{align*}
\end{proof}

\begin{proof}{ of Lemma~\ref{lemma:non-increasing}}
  Without loss of generality, we assume that
  $T = \{ \obj_1, \obj_2, \ldots, \obj_j \}$ and consider
  $T \cup \{\obj_{j+1}\}$.  For brevity, let $i_1..i_2$ (where
  $1 \le i_1 \le i_2 \le n$) denote
  $\{ \obj_i \mid i_1 \le i \le i_2 \}$, i.e., the set of objects with
  indices between $i_1$ and $i_2$ (inclusive).  Let
  $\concat{\cdot}{\cdot}$ denote the concatenation of (column) vectors
  and/or individual values into a (column) vector.
  \begin{align*}
    \EVar{1..j}
   &  = \sum_{\*v \in \*V_{1..j}} %
      \Prob{\*X_{1..j} = \*v} \cdot %
      \Var{ f(\*X) \mid \*X_{1..j} = \*v }\\
    & = \sum_{\*v \in \*V_{1..j}} %
      \Prob{\*X_{1..j} = \*v} \cdot %
      \Big( %
      \Expected{ (f(\*X))^2 \mid \*X_{1..j}=\*v } - (\Expected{ f(\*X) \mid \*X_{1..j}=\*v })^2
      \Big).
  \end{align*}
  We also have
  \begin{align*}
    \EVar{1..j+1}
    & = \sum_{\concat{\*v}{v_{j+1}} \in \*V_{1..j+1}} %
      \Prob{\*X_{1..j+1} = \concat{\*v}{v_{j+1}}} \cdot\\
    & \hspace*{3em}\Big( %
      \Expected{ (f(\*X))^2 \mid \*X_{1..j+1}=\concat{\*v}{v_{j+1}} } - %
      (\Expected{ f(\*X) \mid \*X_{1..j+1}=\concat{\*v}{v_{j+1}} })^2
      \Big)\\
    & = \sum_{\*v \in \*V_{1..j}} \Prob{\*X_{1..j} = \*v} \cdot \Bigg( %
      \sum_{v_{j+1} \in V_{j+1}} %
      \Prob{X_{j+1} = v_{j+1} \mid \*X_{1..j} = \*v} \cdot\\
    &\hspace*{3em} \Big( %
      \Expected{ (f(\*X))^2 \mid \*X_{1..j+1}=\concat{\*v}{v_{j+1}} } - %
      (\Expected{ f(\*X) \mid \*X_{1..j+1}=\concat{\*v}{v_{j+1}} })^2
      \Big)\Bigg)\\
    & = \sum_{\*v \in \*V_{1..j}} \Prob{\*X_{1..j} = \*v} \cdot \Bigg( %
      \sum_{v_{j+1} \in V_{j+1}} %
      \Prob{X_{j+1} = v_{j+1} \mid \*X_{1..j} = \*v} \cdot\\
      &\hspace*{3em} \Expected{ (f(\*X))^2 \mid \*X_{1..j+1}=\concat{\*v}{v_{j+1}} }- %
      \sum_{v_{j+1} \in V_{j+1}} %
      \Prob{X_{j+1} = v_{j+1} \mid \*X_{1..j} = \*v} \cdot\\%
    & \hspace*{6em} (\Expected{ f(\*X) \mid \*X_{1..j+1}=\concat{\*v}{v_{j+1}} })^2 %
      \Bigg)\\
    & = \sum_{\*v \in \*V_{1..j}} \Prob{\*X_{1..j} = \*v} \cdot
    \Bigg( %
      \Expected{ (f(\*X))^2 \mid \*X_{1..j}=\*v }-\\
     & \hspace*{3em}\sum_{v_{j+1} \in V_{j+1}}
      \Prob{X_{j+1} = v_{j+1} \mid \*X_{1..j} = \*v} \cdot%
     (\Expected{ f(\*X) \mid \*X_{1..j+1}=\concat{\*v}{v_{j+1}} })^2 %
      \Bigg).
  \end{align*}
  Comparing the final forms of $\EVar{1..j}$ and $\EVar{1..j+1}$
  above, we see that
  \begin{align*}
    \EVar{1..j} - \EVar{1..j+1}
    & = \sum_{\*v \in \*V_{1..j}} %
      \Prob{\*X_{1..j} = \*v} \cdot (L_1(\*v) - L_0(\*v)),\\
    \text{where}\; L_0(\*v)
    & = (\Expected{ f(\*X) \mid \*X_{1..j}=\*v })^2\\
    \text{and}\; L_1(\*v)
    & =\!\!\!\!\!\sum_{v_{j+1} \in V_{j+1}}\!\!\!\!\!%
      \Prob{X_{j+1} = v_{j+1} \mid \*X_{1..j} = \*v} \cdot\\%
    & \hspace*{3em}(\Expected{ f(\*X) \mid \*X_{1..j+1}=\concat{\*v}{v_{j+1}} })^2.
  \end{align*}
  To show $\EVar{1..j} \ge \EVar{1..j+1}$, it suffices to show that
  $\forall \*v \in \*V_{1..j}: L_1(\*v) \ge L_0(\*v)$.  Given $\*v$, let
  $\eta = \card{V_{j+1}}$ and
  $V_{j+1} = \{ y_1, y_2, \ldots, y_\eta \}$, and for
  $\ell = 1, 2, \ldots, \eta$, let
  \begin{align*}
    p_\ell & = \Prob{X_{j+1} = y_\ell \mid \*X_{1..j}=\*v};\\
    z_\ell & = \Expected{ f(\*X) \mid \*X_{1..j+1}=(\*v, y_\ell) }.
  \end{align*}
  Hence,
  \begin{align*}
    L_0(\*v) = %
    \left(\sum_{\ell=1}^\eta p_\ell z_\ell\right)^2,%
    \;\text{and}\; %
    L_1(\*v) = %
    \sum_{\ell=1}^\eta p_\ell z_\ell^2,%
    \;\text{where}\; %
    \sum_{\ell=1}^\eta p_\ell = 1.
  \end{align*}
  By Lemma~\ref{lemma:power-mean}, $L_1(\*v) \ge L_0(\*v)$.
\end{proof}

\begin{proof}{ of Lemma~\ref{lemma:submodular}}
  Without loss of generality, we assume
  $T = \{ \obj_1, \obj_2, \ldots, \obj_j \}$ and
  $T' = \{ \obj_1, \obj_2, \ldots, \obj_k \}$ where $1 \le j < k < n$.
  As in the proof of Lemma~\ref{lemma:non-increasing}, let $i_1..i_2$
  (where $1 \le i_1 \le i_2 \le n$) denote
  $\{ \obj_i \mid i_1 \le i \le i_2 \}$; furthermore, let
  $i_1..i_2,i_3$ (where $1 \le i_1 \le i_2 < i_3 \le n$) denote
  $i_1..i_2 \cup \{i_3\}$.  Let $\concat{\cdot}{\cdot}$ denote the
  concatenation of (column) vectors and/or individual values into a
  (column) vector.  We shall show that
  \begin{align*}
    J & \le K,\\
    \text{where}\; J & = \EVar{\*X_{1..j}} - \EVar{\*X_{1..j,n}},\\
    \text{and}\; K & = \EVar{\*X_{1..k}} - \EVar{\*X_{1..k,n}}.
  \end{align*}

  Given $\*v \in V_{1..j}$, let $\eta = \card{V_n}$ and
  $V_n = \{ y_1, y_2, \ldots, y_\eta \}$, and for
  $\ell = 1, 2, \ldots, \eta$, let\\
  $p_\ell = \Prob{X_n = y_\ell \mid \*X_{1..j}=\*v} = \Prob{X_n =
    y_\ell}$
  (because $X_i$'s are independent). Following the derivation in
  Lemma~\ref{lemma:non-increasing} (of $\EVar{\*X_{1..j+1}}$ in terms
  of $L_1(\*v)$ and $L_0(\*v)$), we have
  \begin{align*}
    J
    & = \sum_{\*v \in V_{1..j}} \Prob{X_{1..j}=\*v} \cdot\\%
    & \hspace*{3em}\Bigg(%
      \Big(\sum_{\ell=1}^\eta p_\ell\,%
      (\Expected{f(\*X) \mid%
      X_{1..j,n}=\concat{\*v}{y_\ell}})^2\Big)%
      -%
    \Big(\sum_{\ell=1}^\eta p_\ell\,%
      \Expected{f(\*X) \mid%
      X_{1..j,n}=\concat{\*v}{y_\ell}}\Big)^2%
      \Bigg)\\
    & = \sum_{\*v \in V_{1..j}} \Prob{X_{1..j}=\*v} \cdot\\
    & \hspace*{3em}\Bigg(%
      \sum_{1 \le \ell_1 < \ell_2 \le \eta} p_{\ell_1} p_{\ell_2}%
      \Big(%
      \Expected{f(\*X) \mid X_{1..j,n} = \concat{\*v}{y_{\ell_1}}}%
    -
    \Expected{f(\*X) \mid X_{1..j,n} = \concat{\*v}{y_{\ell_2}}}%
      \Big)^2%
      \Bigg),
  \end{align*}
  where the last step follows from Lemma~\ref{lemma:power-mean}.
  Similarly, we have
  \begin{align*}
    K
    & = \sum_{\*v \in V_{1..k}} \Prob{X_{1..k}=\*v} \cdot\\
    & \hspace*{3em}\Bigg(%
      \sum_{1 \le \ell_1 < \ell_2 \le \eta} p_{\ell_1} p_{\ell_2}%
      \Big(%
      \Expected{f(\*X) \mid X_{1..k,n} = \concat{\*v}{y_{\ell_1}}}%
    -%
      \Expected{f(\*X) \mid X_{1..k,n} = \concat{\*v}{y_{\ell_2}}}%
      \Big)^2%
      \Bigg).
  \end{align*}
  Because $X_i$'s are independent, we can rewrite $K$ as follows:
  \begin{align*}
    K
    & = \sum_{\*v \in V_{1..j}} \sum_{\*v' \in V_{j+1..k}} %
      \Prob{X_{1..j}=\*v} \cdot%
      \Prob{X_{j+1..k}=\*v'} \cdot\\
    & \hspace*{3em}\Bigg(%
      \sum_{1 \le \ell_1 < \ell_2 \le \eta} p_{\ell_1} p_{\ell_2}%
      \Big(%
      \Expected{f(\*X) \mid X_{1..k,n} = \concatiii{\*v}{\*v'}{y_{\ell_1}}}
    -%
      \Expected{f(\*X) \mid X_{1..k,n} = \concatiii{\*v}{\*v'}{y_{\ell_2}}}%
      \Big)^2%
      \Bigg)\\
    & = \sum_{\*v \in V_{1..j}} \Prob{X_{1..j}=\*v} \cdot\\
    & \hspace*{3em} \Bigg(%
      \sum_{1 \le \ell_1 < \ell_2 \le \eta} p_{\ell_1} p_{\ell_2} \cdot
      \Bigg(%
      \sum_{\*v' \in V_{j+1..k}} \Prob{X_{j+1..k}=\*v'} \cdot%
    \Big(%
      \Expected{f(\*X) \mid X_{1..k,n} = \concatiii{\*v}{\*v'}{y_{\ell_1}}}\\%
    & \hspace*{6em} -%
      \Expected{f(\*X) \mid X_{1..k,n} = \concatiii{\*v}{\*v'}{y_{\ell_2}}}%
      \Big)^2%
      \Bigg)\Bigg)
  \end{align*}
  Comparing the above with the final form of $J$ derived earlier, we
  see that in order to show $K \ge J$, it suffices to show that for
  all $\*v \in V_{1..j}$ and $1 \le \ell_1 < \ell_2 \le \eta$:
  \begin{align*}    
      &\sum_{\*v' \in V_{j+1..k}} \Prob{X_{j+1..k}=\*v'} \cdot %
      \Big(%
      \Expected{f(\*X) \mid X_{1..k,n} = \concatiii{\*v}{\*v'}{y_{\ell_1}}}%
   -%
      \Expected{f(\*X) \mid X_{1..k,n} = \concatiii{\*v}{\*v'}{y_{\ell_2}}}%
      \Big)^2\\
    & \hspace*{3em} \ge%
      \Big(%
      \Expected{f(\*X) \mid X_{1..j,n} = \concat{\*v}{y_{\ell_1}}}%
      -%
      \Expected{f(\*X) \mid X_{1..j,n} = \concat{\*v}{y_{\ell_2}}}%
      \Big)^2.
  \end{align*}
  Let $\eta' = |V_{j+1..k}|$ and
  $V_{j+1..k} = \{ \*y'_1, \*y'_2, \ldots, \*y'_{\eta'} \}$, and for
  $\ell' = 1, 2, \ldots, \eta'$, let
  \begin{align*}
    p'_{\ell'}
    & = \Prob{\*X_{j+1..k} = \*y_{\ell'}} \;%
      \text{(recalling that $X_i$'s are independent)};\\
    z'_{\ell'}
    & = %
      \Expected{f(\*X) \mid%
      X_{1..k,n} = \concatiii{\*v}{\*y'_{\ell'}}{y_{\ell_1}}}%
      -
     \Expected{f(\*X) \mid%
      X_{1..k,n} = \concatiii{\*v}{\*y'_{\ell'}}{y_{\ell_2}}}.
  \end{align*}
  Then, the inequality to be proven can be rewritten as
  \begin{align*}
    \sum_{\ell'=1}^{\eta'} p'_{\ell'} (z'_{\ell'})^2
    & \ge%
      \Big(%
      \Expected{f(\*X) \mid X_{1..j,n} = \concat{\*v}{y_{\ell_1}}}%
      -%
      \Expected{f(\*X) \mid X_{1..j,n} = \concat{\*v}{y_{\ell_2}}}%
      \Big)^2\\
    & =%
      \Bigg(%
      \Big(\sum_{\ell'=1}^{\eta'} p'_{\ell'} \cdot %
      \Expected{f(\*X) \mid %
      X_{1..k,n} = \concatiii{\*v}{\*y'_{\ell'}}{y_{\ell_1}}}\Big)%
      -\\%
    & \hspace*{3em}\Big(\sum_{\ell'=1}^{\eta'} p'_{\ell'} \cdot %
      \Expected{f(\*X) \mid %
      X_{1..k,n} = \concatiii{\*v}{\*y'_{\ell'}}{y_{\ell_2}}}\Big)%
      \Bigg)^2\\
    & =%
      \Bigg(%
      \sum_{\ell'=1}^{\eta'} p'_{\ell'} \cdot %
      \Big(\Expected{f(\*X) \mid %
      X_{1..k,n} = \concatiii{\*v}{\*y'_{\ell'}}{y_{\ell_1}}}%
      -%
     \Expected{f(\*X) \mid %
      X_{1..k,n} = \concatiii{\*v}{\*y'_{\ell'}}{y_{\ell_2}}}\Big)
      \Bigg)^2\\
    & =%
      \left(\sum_{\ell'=1}^{\eta'} p'_{\ell'} z'_{\ell'} \right)^2,
  \end{align*}
  which follows from Lemma~\ref{lemma:power-mean}.
\end{proof}


\begin{proof}{ of Lemma~\ref{lemma:Compl}}
We start by proving the first part of the lemma.
From the definition, we have that $\EVarComplement{\Objs \setminus \widetilde{T}}=\EVar{\widetilde{T}}$ or $\EVarComplement{\widetilde{T}}=\EVar{\Objs \setminus \widetilde{T}}$, for any $\widetilde{T}\subseteq \Objs$.
Hence, if $\overline{T}$ is a solution for the \MinVarComplement\ problem such that $\EVarComplement{\overline{T}}\leq \alpha \EVarComplement{\overline{T}^*}$ and $\sum_{\obj_i\in \overline{T}}c_i\geq \left(\sum_{\obj_i \in \Objs}\right) c_i- C$, where $\overline{T}^*$ is the optimum set, then $T=\Objs\setminus \overline{T}$ is a solution for the \MinVar\ problem such that $\EVar{T} \leq \alpha \EVar{\Objs\setminus \overline{T}^*}$
and $\sum_{\obj_i\in T}c_i\leq C$. From our first observation notice that $\Objs\setminus \overline{T}^*$ is the optimum set to clean for the \MinVar\ problem.
The same holds in the other direction: An $\alpha$-approximation solution for the \MinVar\ problem is also an $\alpha$-approximation for the \MinVarComplement\ problem. Hence, \MinVar\ problem can be mapped to \MinVarComplement\ problem.

Next we show the second part of the lemma.

Let $T_1\subseteq T_2\subseteq \Objs$. We have
$$\EVarComplement{T_1}=\EVar{\Objs\setminus T_1}\leq \EVar{\Objs\setminus T_2}=\EVarComplement{T_2},$$
because $\Objs\setminus T_1\supseteq \Objs\setminus T_2$ and $\EVar{\cdot}$ is monotone non-increasing so $\EVarComplement{\cdot}$ is monotone non-decreasing.

Now we prove the function is submodular.
Let $T_1\subset T_2\subseteq \Objs$ and $x\notin T_2$ for $x\in \Objs$.
Notice that $\Objs\setminus T_1\supset \Objs\setminus T_2$.
Let $T_1'=\Objs\setminus (T_1\cup \{x\})$ and $T_2'=\Objs\setminus (T_2\cup \{x\})$ with $T_1'\supset T_2'$.
Finally, recall that $\EVar{\cdot}$ is submodular.
We have,
\begin{align*}
\EVarComplement{T_1\cup \{x\}}-\EVarComplement{T_1}
&=\EVar{\Objs\setminus (T_1\cup \{x\})}-\EVar{\Objs\setminus T_1}
=\EVar{T_1'}-\EVar{T_1'\cup \{x\}}\\
&=-(\EVar{T_1'\cup \{x\}}-\EVar{T_1'})
\geq -(\EVar{T_2'\cup \{x\}}-\EVar{T_2'})\\
&=\EVar{T_2'}-\EVar{T_2'\cup \{x\}}
=\EVarComplement{T_2\cup \{x\}}-\EVarComplement{T_2}.
\end{align*}
\end{proof}

\subsection{Missing proofs for application in fact checking}
\label{appndx:ApplFactChecking}

\begin{proof}{ of Theorem~\ref{th:runtimeDepClaims}}
We first consider the Uniqueness($\duplicity{q^\circ(\*u)}{\*X}$) and later we extend for Fairness ($\bias{q^\circ(\*u)}{\*X}$) and Robustness ($\fragility{q^\circ(\*u)}{\*X}$).
Let $t$ be an instance of $T$.
We have that $$\EVar{T}=\sum_{t\in T}\Prob{t}(\sum_{q\in Q} \Var{\mathbf{1}[\Delta(q, q^\circ) \ge 0] \mid t} + \sum_{q, q'\in Q}\Cov{\mathbf{1}[\Delta(q, q^\circ) \ge 0], \mathbf{1}[\Delta(q', q^\circ) \ge 0] \mid t}).$$
Let assume a claim $q\in Q$ and let $T_q$ be the objects in $T\cap (q\cup q^\circ)$.
We observe that $$\sum_{t\in T}\Prob{t}\Var{\mathbf{1}[\Delta(q, q^\circ) \ge 0]\mid t}=\\\sum_{t\in T_q}\Prob{t}\Var{\mathbf{1}[\Delta(q, q^\circ) \ge 0]\mid t}$$ since the different values of the tuples that do not belong in $q$ or $q^\circ$ do not change the variance of $\mathbf{1}[\Delta(q, q^\circ) \ge 0]$. Equivalently, let $q, q'\in Q$ and let $T_{qq'}$ be the objects in $T\cap (q\cup q'\cup q^\circ)$.
We observe that
\begin{align*}
\sum_{t\in T}\Prob{t}\Cov{\mathbf{1}[\Delta(q, q^\circ) \ge 0],\mathbf{1}[\Delta(q', q^\circ) \ge 0] \mid t}&=\\
&\hspace*{-3em}\sum_{t\in T_{qq'}}\Prob{t}\Cov{\mathbf{1}[\Delta(q, q^\circ) \ge 0],\mathbf{1}[\Delta(q', q^\circ) \ge 0] \mid t}
\end{align*}
since the different values of the tuples that do not belong in $q, q'$ or $q^\circ$ do not change the covariance of $\mathbf{1}[\Delta(q, q^\circ) \ge 0],\mathbf{1}[\Delta(q', q^\circ) \ge 0]$.

Based on the observations above we can give a polynomial time algorithm to compute the $\EVar{T}$.
We first start computing the variances. For each claim $q\in Q$ we sum over all the instances of $T_q$, which are $O(V^b)$ for a value $b\leq 2W$
(notice that $|q\cup q^\circ|\leq 2W$).
For each such instance $t$ it remains to compute $\Var{\mathbf{1}[\Delta(q, q^\circ) \ge 0]\mid t}$. This can be computed in $O(V^{2W-b}W)$ time, by creating the distribution of $\mathbf{1}[\Delta(q, q^\circ) \ge 0]$, so in total we need $O(mV^{2W}W+n)$ to find all the variances of all claims.
Then we continue with computing the covariances.
For a pair of claims $q, q'\in Q$ we sum over all the instances of $T_{qq'}$, which are $O(V^b)$ for a value $b\leq 3W$
(notice that $|q\cup q'\cup q^\circ|\leq 3W$).
For each such instance $t$ it remains to compute $\Cov{\mathbf{1}[\Delta(q, q^\circ) \ge 0],\mathbf{1}[\Delta(q', q^\circ) \ge 0] \mid t}$, which can be computed in $O(V^{3W-b}W)$, by creating the distribution of $\mathbf{1}[\Delta(q, q^\circ) \ge 0]\cdot \mathbf{1}[\Delta(q', q^\circ) \ge 0]$, so in total we need $O(m^2V^{3W}W+n)$ to find all covariances of all pairs of claims.

If we consider the Robustness or Fairness then the only difference is that we compute\\
$\Var{s_k \cdot (\min\{\Delta(q_k,q^\circ), 0\})^2\mid t}$ (or $\Var{ s_k \cdot \Delta(q_k(\*X), q^\circ(\*u))}$) and\\
$\Cov{s_k \cdot (\min\{\Delta(q_k,q^\circ), 0\})^2, s_{k'} \cdot (\min\{\Delta(q_{k'},q^\circ), 0\})^2 \mid t}$ (or \\
$\Cov{s_k \cdot \Delta(q_k(\*X), q^\circ(\*u)),s_{k'} \cdot \Delta(q_{k'}(\*X), q^\circ(\*u))}$).
The first quantities can be computed in $O(V^{2W}W)$ and the second ones in $O(V^{3W}W)$.

Overall, we conclude that the $\EVar{T}$ of $\bias{q^\circ(\*u)}{\*X}$, $\duplicity{q^\circ(\*u)}{\*X}$, and $\fragility{q^\circ(\*u)}{\*X}$
can be computed in $O(m^2V^{3W}W+n)$ time in  the worst case.
\end{proof}

\begin{proof}{ of Theorem~\ref{theorem:minvar-maxpr-equivalence}}
We first start by showing that if the independence assumption holds then the result follows easily by Lemma~\ref{lemma:linear-modular}.
In the first part of Lemma~\ref{lemma:linear-modular} we showed that the \MinVar\ problem is equivalent to maximizing $\sum_{\obj_i \in T} a_i^2 \Var{X_i}$.
In the second part of the Lemma, we showed that \MaxPr\ is equivalent to maximizing $\sum_{\obj_i \in T} a_i^2 \sigma_i^2$ and the result follows easily.

We can extend the above argument when the independence assumption does not hold.
From the beginning of Lemma~\ref{lemma:linear-modular}, the objective of \MinVar\ problem is
$$
    \min_{T\subseteq \Objs} \sum_{\obj_i \in \Objs \setminus T}%
      \,\sum_{\obj_j \in \Objs \setminus T}%
      \Cov{a_iX_i, a_jX_j}
    =\max_{T\in \subseteq \Objs} \sum_{\obj_i \in T}%
      \,\sum_{\obj_j \in  T}%
      \Cov{a_iX_i, a_jX_j}.
$$
From the second part of Lemma~\ref{lemma:linear-modular}, the objective of \MaxPr\ problem is
\begin{align*}
\max_{T\subseteq \Objs} \Prob{f(\*X) < f(\*u) - \tau %
     \mid \*X_{\Objs \setminus T} = \*u_{\Objs \setminus T}} %
   & = \max_{T\subseteq \Objs} \Prob{N(0, \sum_{\obj_i \in T}%
      \,\sum_{\obj_j \in  T}%
      \Cov{a_iX_i, a_jX_j}) < -\tau}\\ %
   & = \max_{T\subseteq \Objs}\Prob{N(0, 1) < %
    {-\tau \over \sqrt{\sum_{\obj_i \in T}%
      \,\sum_{\obj_j \in  T}%
      \Cov{a_iX_i, a_jX_j}}}}.
\end{align*}
Since $\tau>0$, this is equivalent to
$$\max_{T\subseteq \Objs} \sum_{\obj_i \in T}%
      \,\sum_{\obj_j \in  T}%
      \Cov{a_iX_i, a_jX_j},$$
and the result follows.
\end{proof}

\remove{

\subsection{Additional figures}
Figures~\ref{appndx:fig:expr:uniqueness:univ-unic}, \ref{appndx:fig:expr:uniqueness:logv-unic}, \ref{appndx:fig:expr:uniqueness:nonv-unic} show the effectiveness of various algorithms in reducing the uncertainty in the uniqueness of a claim for value distributions that are generated using $\mathsf{UR}_x$, $\mathsf{LN}_x$, $\mathsf{SM}_x$, respectively..

\label{appndx:misFig}
\begin{figure*}[h]
\begin{subfigure}{0.33\linewidth}
  \input{figs/expr/perc/univ-unic-r50}
  \caption{$\Gamma=50$}
\end{subfigure}\hfill
\begin{subfigure}{0.33\linewidth}
  \input{figs/expr/perc/univ-unic-r100}
  \caption{$\Gamma=100$}
\end{subfigure}
\begin{subfigure}{0.33\linewidth}
  \input{figs/expr/perc/univ-unic-r150}
  \caption{$\Gamma=150$}
\end{subfigure}
\begin{subfigure}{0.33\linewidth}
  \input{figs/expr/perc/univ-unic-r200}
  \caption{$\Gamma=200$}
\end{subfigure}\hfill
\begin{subfigure}{0.33\linewidth}
  \input{figs/expr/perc/univ-unic-r250}
  \caption{$\Gamma=250$}
\end{subfigure}
\begin{subfigure}{0.33\linewidth}
  \input{figs/expr/perc/univ-unic-r300}
  \caption{$\Gamma=300$}
\end{subfigure}
\caption{Effectiveness of various algorithms in reducing the uncertainty in the uniqueness of a claim asserting an aggregate result to be as small as
    $\Gamma$. Value distributions are generated using $\mathsf{UR}_x$.}
\label{appndx:fig:expr:uniqueness:univ-unic}
\end{figure*}

\begin{figure*}[h]
\begin{subfigure}{0.33\linewidth}
  \input{figs/expr/perc/logv-unic-r30}
  \caption{$\Gamma=3.0$}
\end{subfigure}\hfill
\begin{subfigure}{0.33\linewidth}
  \input{figs/expr/perc/logv-unic-r35}
  \caption{$\Gamma=3.5$}
\end{subfigure}
\begin{subfigure}{0.33\linewidth}
  \input{figs/expr/perc/logv-unic-r40}
  \caption{$\Gamma=4.0$}
\end{subfigure}
\begin{subfigure}{0.33\linewidth}
  \input{figs/expr/perc/logv-unic-r45}
  \caption{$\Gamma=4.5$}
\end{subfigure}\hfill
\begin{subfigure}{0.33\linewidth}
  \input{figs/expr/perc/logv-unic-r50}
  \caption{$\Gamma=5.0$}
\end{subfigure}
\begin{subfigure}{0.33\linewidth}
  \input{figs/expr/perc/logv-unic-r55}
  \caption{$\Gamma=5.5$}
\end{subfigure}
\caption{Effectiveness of
    various algorithms in reducing the uncertainty in the uniqueness
    of a claim asserting an aggregate result to be as small as
    $\Gamma$. Value distributions are generated using
    $\mathsf{LN}_x$.}
\label{appndx:fig:expr:uniqueness:logv-unic}
\end{figure*}

\begin{figure*}[h]
\begin{subfigure}{0.33\linewidth}
  \input{figs/expr/perc/nonv-unic-r50}
  \caption{$\Gamma=50$}
\end{subfigure}\hfill
\begin{subfigure}{0.33\linewidth}
  \input{figs/expr/perc/nonv-unic-r100}
  \caption{$\Gamma=100$}
\end{subfigure}
\begin{subfigure}{0.33\linewidth}
  \input{figs/expr/perc/nonv-unic-r150}
  \caption{$\Gamma=150$}
\end{subfigure}
\begin{subfigure}{0.33\linewidth}
  \input{figs/expr/perc/nonv-unic-r200}
  \caption{$\Gamma=200$}
\end{subfigure}\hfill
\begin{subfigure}{0.33\linewidth}
  \input{figs/expr/perc/nonv-unic-r250}
  \caption{$\Gamma=250$}
\end{subfigure}
\begin{subfigure}{0.33\linewidth}
  \input{figs/expr/perc/nonv-unic-r300}
  \caption{$\Gamma=300$}
\end{subfigure}
\caption{Effectiveness of
    various algorithms in reducing the uncertainty in the uniqueness
    of a claim asserting an aggregate result to be as small as
    $\Gamma$. Value distributions are generated using
    $\mathsf{SM}_x$.}
\label{appndx:fig:expr:uniqueness:nonv-unic}
\end{figure*}

\subsection{Absolute improvement}
\label{appndx:diff}
\begin{figure*}[h]
\begin{subfigure}{0.49\linewidth}
  \input{figs/expr/perc/univ-unic-abs-improvement}
  \caption{$\mathsf{UR}_x$}
\end{subfigure}\hfill
\begin{subfigure}{0.49\linewidth}
  \input{figs/expr/perc/logv-unic-abs-improvement}
  \caption{$\mathsf{LN}_x$}
\end{subfigure}
\caption{Absolute
    improvement of \Greedy\MinVar\ over \GreedyNaive\ in reducing
    uncertainty, for the same scenarios in
    Figures~\ref{appndx:fig:expr:uniqueness:univ-unic}
    and~\ref{appndx:fig:expr:uniqueness:logv-unic}, respectively.}
\label{fig:expr:uniqueness:abs-improvement}
\end{figure*}

While it may appear from these figures that the differences among
algorithms are small when the initial uncertainty is high, we note
that if we instead examine the improvement in uncertainty by
\Greedy\MinVar\ over \GreedyNaive\ in absolute terms, we would in fact
see bigger improvements in cases with larger initial uncertainty.
Figure~\ref{fig:expr:uniqueness:abs-improvement} shows the absolute
improvement (in the amount of expected variance reduced) of
\Greedy\MinVar\ over \GreedyNaive, for the same scenarios in
Figures~\ref{appndx:fig:expr:uniqueness:univ-unic} ($\mathsf{UR}_x$)
and~\ref{appndx:fig:expr:uniqueness:logv-unic} ($\mathsf{LN}_x$); the
scenario of Figures~\ref{appndx:fig:expr:uniqueness:nonv-unic}
($\mathsf{SM}_x$) is similar.  Each curve shows, for a specific value
of $\Gamma$, the improvement as a function of the budget.  The legends
list the $\Gamma$ values in descending order of the corresponding
initial uncertainties.  We see that this ordering is fairly consistent
with the ordering of the curves; i.e., a higher initial uncertainty
translates to a bigger absolute improvement of \Greedy\MinVar\ over
\GreedyNaive.  For example, in
Figure~\ref{fig:expr:uniqueness:abs-improvement}b, improvement is the
biggest for $\Gamma=4.0$, which has the peak initial uncertainty;
improvement is also big for $\Gamma=3.5$ and $\Gamma=4.5$, whose
initial uncertainties are the next highest.  In contrast, the absolute
improvement for $\Gamma=3$ is small, even though the relative
improvement shown in Figure~\ref{appndx:fig:expr:uniqueness:logv-unic}a is
huge.
The results in Figure~\ref{fig:expr:uniqueness:abs-improvement}a
for $\mathsf{UR}_x$  are not as clear as for $\mathsf{LN}_x$. However
we can observe that for $\Gamma=50$ the absolute improvement is very small,
while the relative improvement was huge as shown in Figure~\ref{appndx:fig:expr:uniqueness:univ-unic}b.

Figure~\ref{fig:expr:uniqueness:abs-improvement} also allows us to see
the effect of the budget constraint on the improvement of
\Greedy\MinVar\ over \GreedyNaive.  When the budget becomes either
very tight or very generous, the difference between \Greedy\MinVar\
over \GreedyNaive\ becomes smaller.  This effect is consistent with
intuition: a tight budget means limited options, while a generous
budget means choices matter less since most uncertainty will be
removed anyway.

\subsection{Effectiveness in action}
\label{appndx:counter}
\begin{figure*}
\begin{subfigure}{0.49\linewidth}
  \input{figs/expr/perc/uni-unic-r100-EM}
  \caption{Expected uniqueness}
\end{subfigure}\hfill
\begin{subfigure}{0.49\linewidth}
  \input{figs/expr/perc/uni-unic-r100-Var}
  \caption{Variance of uniqueness}
\end{subfigure}
\caption{Expectation and variance of claim uniqueness
    as functions of budget for a specific scenario with value
    distributions from $\mathsf{UR}_x$ and $\Gamma=100$.}
\label{fig:Var}
\end{figure*}
In Figures~\ref{fig:Var}a and \ref{fig:Var}b we plot the expectation and the variance of claim uniqueness resulted from each algorithm's decision as a function of budget. Knowing that the true uniqueness is $1$, we can observe that $\mathsf{Best}$ and \Greedy\MinVar\ perform much better than \GreedyNaive.

For the second experiment,
we use
the same data and workload as in subsection ``Effectiveness in Action'' in Section~\ref{sec:expr},
but in addition to the hidden
true values of the objects, we also randomly draw their current
(noisy) values.  For the specific scenario we experiment with, there
is no counter to the original claim assuming current object values;
however, with true object values, there indeed exists a strong counter
to the original claim.  In other words, a fact-checker must clean some
values in order to find a counter.  We compare how \Greedy\MaxPr\ and
\GreedyNaive\ help with this task, as we vary the available budget.
After cleaning only $3$ values and using $8\%$ of the total cost,
\Greedy\MaxPr\ is able to find a counter to the original claim.  In
comparison, \GreedyNaive\ is much less effective: it finds a counter
only after cleaning $15$ values and using $21\%$ of the total cost.
The result here of course is for one specific scenario, but it does
reaffirm the effectiveness of \Greedy\MaxPr\ in maximizing the
probability of finding a counter.

\subsection{Efficiency of Algorithms}
\label{appndx:Ef}
\begin{figure*}[h]
\begin{subfigure}{0.49\linewidth}
  \begin{tikzpicture}[yscale=0.5, xscale=0.6]
\begin{axis}[
    y label style={at={(axis description cs:0.1,.5)},anchor=south},
    xlabel=\emph{Budget},
    ylabel=\emph{{\large Time (sec)}},
    xmin=0.9, xmax=5.1,
    ymin=0, ymax=110,
    xtick={1, 2, 3, 4, 5},
    xticklabels={1\%, 5\%, 10\%, 20\%, 30\%},
    legend pos=north west,
    ymajorgrids=true,
    grid style=dashed,
]

\addplot[
    color=blue,
    mark=*,
    ]
    coordinates {
    (1, 8.737) (2, 28.057) (3, 48.02) (4, 80.498) (5, 108.186)
    };
     \addlegendentry{GreedyMinVar};
\end{axis}
\end{tikzpicture} 
  \caption{$n = 10{,}000$, varying budget}
\end{subfigure}\hfill
\begin{subfigure}{0.49\linewidth}
  \begin{tikzpicture}[yscale=0.5, xscale=0.6]
\begin{axis}[
    y label style={at={(axis description cs:0.15,.5)},anchor=south},
    xlabel=\emph{$n$},
    ylabel=\emph{{\large $\log_{10}$ Time (sec)}},
    xmin=0.9, xmax=5.1,
    ymin=1, ymax=4.5,
    xtick={1, 2, 3, 4, 5},
    xticklabels={5000, 10000, 100000, 500000, 1000000},
    legend pos=north west,
    ymajorgrids=true,
    grid style=dashed,
]

\addplot[
    color=blue,
    mark=*,
    ]
    coordinates {
     (1, 1.23) (2, 1.59) (3, 2.86) (4, 3.77) (5, 4.12)
    };
     \addlegendentry{GreedyMinVar};
\end{axis}
\end{tikzpicture} 
  \caption{varying $n$, budget $C=5000$}
\end{subfigure}
\caption{Running time of
    \Greedy\MinVar\ when reducing uncertainty in uniqueness, for the
    same scenario in Figure~\ref{fig:expr:uniqueness:All}a
    ($\Gamma=100$).}
\label{fig:expr:time:univ-unic}
\end{figure*}
In
Figure~\ref{fig:expr:time:univ-unic}a, we show the running time of
\Greedy\MinVar\ as we give it increasingly larger budgets to work
with.  We see that running time increases roughly linearly with
budget.  Even with a budget that allows $30\%$ of all data to be
cleaned, \Greedy\MinVar\ completes under $2$ minutes.
In Figure~\ref{fig:expr:time:univ-unic}b, we
show how \Greedy\MinVar's running time (in $\log_{10}$ scale) increases with the size of the
dataset.
We can observe that each time that the size of the data set increases by a factor of $10$,
the running time to clean about $1{,}000$ tuples is $18$--$19$ times larger.
Even with a large dataset containing $100{,}000$ uncertain
values, it takes less than $12$ minutes for \Greedy\MinVar\ to
suggest cleaning $1{,}000$ values, which translates to about $0.725$
seconds per recommendation.  Considering that data cleaning is
typically labor-intensive, we believe this overhead of \Greedy\MinVar\
is negligible in practice.

Finally, we note that \Greedy\MaxPr\ and \Greedy\MinVar\ have similar
computation cost, so the conclusions in this experiment also apply to
\Greedy\MaxPr.

\subsection{Ascertaining Claim Quality vs.\ Finding Counters}
\label{appndx:vs}
Figure~\ref{fig:expr:diff-obj}a compares how the two algorithms
achieve the objective of ascertaining claim fairness, while
Figure~\ref{fig:expr:diff-obj}b compares how they achieve the
objective of maximizing the chance of finding counters.  The vertical
axes show the respective objective function values, which are computed
given the choices made by the algorithms.  Note that the dataset's
current values do not affect uncertainty in claim fairness, but do
affect the chance of finding counters significantly.  Hence, for
Figure~\ref{fig:expr:diff-obj}b, we repeat each experiment $100$ times
with different random draws of the current values, and report the
average.

From Figure~\ref{fig:expr:diff-obj}, we see that, as expected, each
algorithm does well in terms of its intended objective.  What is more
revealing is how poorly they do with regard to the other objective.
While $\mathsf{Optimum}$ quite effectively reduces the uncertainty in
claim fairness (Figure~\ref{fig:expr:diff-obj}a), it does not offer a
good chance of finding counters even when given generous budgets
(Figure~\ref{fig:expr:diff-obj}b).  On the other hand, \Greedy\MaxPr\
quickly increases the chance of finding counters
(Figure~\ref{fig:expr:diff-obj}b), but it is much less helpful in
ascertaining claim fairness (Figure~\ref{fig:expr:diff-obj}a).  In
fact, when given a budget above $48\%$ of the total cost of cleaning all data,
\Greedy\MaxPr\ simply refuses
to clean any more values because doing so would actually decrease the
chance of finding a counter (which explains why its achieved objective
function values stay flat beyond this point).  This questionable
behavior illustrates the danger of following a purely utilitarian
approach of cleaning data for fact-checking that just seeks to
maximize the chance of countering a claim.

\begin{figure*}
\begin{subfigure}{0.49\linewidth}
  \input{figs/expr/perc/diff-obj-var}
  \caption{uncertainty in fairness}
\end{subfigure}\hfill
\begin{subfigure}{0.49\linewidth}
  \input{figs/expr/perc/diff-obj-pr}
  \caption{chance of finding counters}
\end{subfigure}
\caption{Comparison between
    $\mathsf{Optimum}$ (for \MinVar) and \Greedy\MaxPr\ (for \MaxPr)
    in how they achieve two different objectives.}
\label{fig:expr:diff-obj}
\end{figure*}
} 

\end{document}